\documentclass[a4paper,10pt]{article}
\usepackage[utf8x]{inputenc}
\usepackage{graphicx}
\usepackage{amssymb,amsmath,color,cite}

\begin{document}
\title{Infinite square-well, trigonometric P\"oschl-Teller and other potential wells with a moving barrier}
\author{Alonso Contreras-Astorga \\
C\'atedras CONACYT -- Departamento de F\'isica, Cinvestav, A.P. 14-740, 07000 \\Ciudad de M\'exico, Mexico \\
Department of Physics, Indiana University Northwest, 3400 Broadway, Gary IN 46408, USA  \\ alonso.contreras@conacyt.mx ; alonso.contreras.astorga@gmail.com 
\and V\'eronique Hussin \\
Centre de Recherches Math\'ematiques \& Department de Math\'ematiques et de Statistique, \\
Universit\'e de Montr\'eal, Montr\'eal QC H3C 3J7, Canada \\
veronique.hussin@umontreal.ca}

\date{}

\maketitle

\begin{abstract}
Using mainly two techniques, a point transformation and  a time dependent supersymmetry, we construct in sequence several quantum infinite potential wells with a moving barrier. We depart from the well known system of a one-dimensional particle in a box.  With a point transformation, an infinite square-well potential with a moving barrier is generated. Using time dependent supersymmetry, the latter leads to a trigonometric P\"oschl-Teller potential with a moving barrier. Finally, a confluent time dependent supersymmetry transformation is implemented to generate new infinite potential wells, all of them with a moving barrier. For all systems, solutions of the corresponding time dependent Schr\"odinger equation fulfilling boundary conditions are presented in a closed form.

\end{abstract}

\section{Introduction}
\label{intro}

There are physical problems where the boundary conditions of the underlying equation can move. Examples of them are the so called Stefan problems, where temperature as a function of position and time on a system of water and ice has to be found, the interface water-ice imposes a boundary condition that changes its position with time \cite{Crank}. Another example was drafted by Fermi, he theorized the origin of cosmic radiation as particles accelerated by collisions with a moving magnetic field \cite{Fermi}, this problem was later studied by Ulam \cite{Ulam} in a classical framework where the statistical properties of particles in a box with oscillating infinite barriers were analyzed numerically.  In this paper, we are interested in systems ruled by the time dependent Schr\"odinger equation. In particular, we show how different quantum systems with a moving boundary condition and their solutions can be generated using basically two tools, a point transformation and a time dependent supersymmetry. 

The point transformation we use was introduced in \cite{Ray82,Bluman83} where the authors mapped solutions between two time dependent Schr\"odinger equations with different potentials. This transformation can be used, for example, to map solutions of the harmonic oscillator to solutions of the free particle system. 

On the other hand, the supersymmetry technique or SUSY helps as well to map solutions between two Schr\"odinger equations, but in this case the potentials share properties like asymtotic behavior or a similiar discrete spectrum in case of time-independent potentials. A first version links two time independent one-dimensional Schr\"odinger equations \cite{Cooper95,Fernandez10}. In this article we use the time dependent version, that links two time dependent Schr\"odinger equations \cite{Matveev,Bagrov95}. The involved potentials are referred as SUSY partners and if the link is made through a first-order differential operator, often called intertwining operator, the  technique is known as 1-SUSY. Examples of time dependent SUSY partners of the harmonic oscillator can be found in \cite{Zelaya17,contreras17}. 

F. Finkel et.al. \cite{Finkel99} showed that the time independent SUSY technique, and the time dependent version were related by the previously mentioned point transformation.

The structure of this article is as follows. The quantum particle in a box is revised in Sec. \ref{Sec 1}.  In Sec. \ref{Sec 2}, we  use a point transformation to generate an infinite square-well potential with a moving barrier. A brief review of time dependent SUSY is given in Sec. \ref{Sec 3} and it is applied to the infinite square-well potential with a moving barrier to generate the exactly solvable system of a P\"oschl-Teller potential with a moving barrier. In Sec. \ref{Sec 4}, we apply for the second time a supersymmetric transformation to the infinite square-well potential to obtain a biparametric family of infinite potential wells with a moving barrier. Exact solutions of the time dependent Schr\"odinger equation for each potential are given in the corresponding section. 
We finish this article with our conclusion.

\section{Quantum infinite square-well potential} \label{sec: Infinite potential square well} \label{Sec 1}

The quantum particle in a one-dimensional infinite square-well potential or particle in a box is a common example of an exact solvable model in textbooks, see for example \cite{Cohen,Shankar,Landau}. It represents a particle trapped in the interval $0 < y < L$ with  impenetrable barriers, placed at zero and $L$, and inside that one-dimensional box the particle is free to move. The corresponding time independent Schr\"odinger equation is 
\begin{eqnarray}
\frac{d^2}{dy^2}\psi(y) +  \frac{2m}{\hbar^2} \left( E- \widetilde{V}_0 \right) \psi(y)=0, 
\end{eqnarray}
where $E$ is a real parameter representing the energy of the particle, $m$ is its mass and $\hbar$ is Planck's constant. Through this article we will use units where $m=1/2$ and $\hbar=1$. The one-dimensional infinite potential well $\widetilde{V}_0(y)$ is: 
\begin{eqnarray}
\widetilde{V}_0(y) = \left\{ \begin{array}{cc} 
                0, & \quad 0 < y < L, \\
                \infty, & \quad \text{otherwise}, 
                \end{array} \right. \label{Vy}
\end{eqnarray}
where $L$ is a positive real constant. The solution of this eigenvalue problem is well known, eigenfunctions and eigenvalues are given by   
\begin{eqnarray}
\psi_n(y)= \sqrt{\frac{2}{L}} \sin\left(\frac{n \pi y}{L} \right) , \qquad E_n = \left(\frac{n \pi}{L} \right)^2, \qquad n=1,2,3,\dots.  \label{Inf eigenfunctions}
\end{eqnarray}
Functions $\psi_n(y)$ satisfy the boundary conditions $\psi_n(0)=\psi_n(L)=0$. We will use this system to construct a variety of infinite potential wells where one of the barriers is moving.

\section{From the particle in a box to the infinite square-well potential with a moving barrier} \label{Sec 2}

In this section we will use a point transformation in order to obtain from the stationary potential (\ref{Vy}) an infinite  square-well potential with a moving barrier. First we introduce a general point transformation \cite{Ray82,Bluman83,Finkel99,Schulze14} and then we apply it on the particle in a box system. The simplified notation for the transformation was introduced in \cite{Schulze14}.   

\subsection{Point transformation} \label{sec: Point Transformation}

Consider a one dimension time independent Schr\"odinger equation in the spatial variable $y$ as 
\begin{eqnarray}
\frac{d^2}{dy^2}\psi(y) + \left(E- \widetilde{V}_0\right)\psi(y)=0  \label{Finkel TISE}
\end{eqnarray}
where $\widetilde{V}_0=\widetilde{V}_0(y)$ and a solution $\psi$ are known. Now let us take arbitrary functions $A=A(t)$ and $B=B(t)$ and let the variable $y$ be defined in terms of a temporal parameter $t$ and a new spatial variable $x$ as:
\begin{eqnarray}
y(x,t)= x \exp\left[ 4 \int A(t) dt   \right] + 2 \int B(t) \exp\left[ 4 \int A(t) dt   \right] dt \label{FinkelVariable} 
\end{eqnarray}
then the function
\begin{eqnarray}
\phi(x,t)&=& \psi(y(x,t)) \exp \left\{-i \left[ A(t)x^2 + B(t)x + E \int \exp\left[8 \int A(t) dt \right] dt     \right.  \right. \nonumber \\
& & \left. \left.  + \int \left[2 i A(t)+B^2(t) \right]dt  \right]    \right\}, \label{FinkelFunction}
 \end{eqnarray}
is solution of the equation
\begin{eqnarray}
i \frac{\partial}{\partial t} \phi(x,t) + \frac{\partial^2}{\partial x^2} \phi(x,t) - V_0(x,t) \phi(x,t)=0. \label{Time SE}
\end{eqnarray} 
The last equation is a time dependent Schr\"odinger equation where the potential is given by 
\begin{eqnarray}
V_0(x,t)&=& \widetilde{V}_0(y(x,t)) \exp \left[8 \int A(t)dt  \right]+ \left[ \frac{d}{dt}A(t)-4 A^2(t) \right] x^2 \nonumber \\
& & + \left[ \frac{d}{dt}B(t) - 4 A(t) B(t)  \right] x. \label{FPot}
\end{eqnarray}

\subsection{Infinite square-well potential with a moving barrier}

We can use the point transformation to obtain an infinite square-well potential with a moving barrier. Without considering for this moment boundary conditions of the problem, we will transform the potential $\widetilde{V}_0(y)=0$ into $V_0(x,t)=0$. Apparently we are mapping a potential to itself but it will not be the case once we incorporate the boundary conditions. To make this transformation, functions $A(t)$ and $B(t)$ such that $V_0(x,t)= \widetilde{V}_0(y)=0$ in  (\ref{FPot})  need to be found. By setting $\widetilde{V}_0(y)=0$ and $V_0(x,t)=0$ in \eqref{FPot} we get  
\begin{eqnarray}
0&=&  \left[ \frac{d}{dt}A(t)-4 A^2(t) \right] x^2  + \left[ \frac{d}{dt}B(t) - 4 A(t) B(t)  \right] x. \label{FinkelPotential}
\end{eqnarray}
Coefficients of the previous polynomial in $x$ give us a system of coupled differential equations that can be solve:  
\begin{align}
&\frac{d}{dt}A(t)-4 A^2(t)=0, &\quad \Rightarrow \quad & A(t)=- \frac{1}{4t +c_1}; \nonumber \\
& \frac{d}{dt}B(t) - 4 A(t) B(t) = 0, &\quad \Rightarrow \quad & B(t)=  \frac{c_2}{4t + c_1};  \label{ABC}
\end{align}
where $c_1$ and $c_2$ are real constants. Once these two functions are known, the change of variable defined in \eqref{FinkelVariable} can be evaluated, 
\begin{eqnarray}
y(x,t)= \frac{2 x - c_2}{2(4t + c_1)}. \label{yvalue}
\end{eqnarray}
At this point, we can discuss boundary conditions of the potential $V_0(x,t)$. The barriers of the potential (\ref{Vy}) of the initial problem are located at $y_1=0$ and at $y_2 = L$, using the change of variable \eqref{yvalue} the new barriers will then be placed at $x_1= c_2/2$ and $x_2= 4 L t + c_1 L + c_2/2$, respectively. Thus, the potential $V_0(x,t)$ is  
\begin{align}
V_0(x,t) = \left\{ \begin{array}{lll} 
                0, & \quad & x_1 < x < \ell(t), \\
                \infty, & \quad & \text{otherwise}, 
                \end{array} \right. \label{square moving}
\end{align} 
where 
\begin{equation}
\ell(t)= 4 L t + c_1 L + c_2/2.
\end{equation}
This potential is an infinite square-well potential with a moving barrier. The meaning of the constants $c_1$ and $c_2$ can be extracted directly from the position of the boundaries of this potential. Indeed, the position of the fixed barrier is $c_2/2$, while the moving barrier is located at $\ell(t)$ and it is moving with a constant velocity $4L$. At $t_0=-c_1/4$ we have an ill defined problem (a particle in a box of length zero), so we should avoid this singularity, moreover, this time $t_0$ separates two problems: one of a contracting box and one where the potential well is expanding as time increases. 

Finally, solutions of the time dependent Schr\"odinger equation \eqref{Time SE} where $V_0(x,t)$ is \eqref{square moving} can be constructed using  \eqref{Inf eigenfunctions}, \eqref{FinkelFunction}, \eqref{ABC} and \eqref{yvalue}:
\begin{eqnarray}
\phi_n(x,t)&=& \sqrt{\frac{2}{L(4t+c_1)}} \sin \left\{ \frac{n \pi}{L} \left[ \frac{2x - c_2}{2(4t + c_1)}  \right]  \right\} \exp \left[ \frac{i}{4t+c_1} x^2- \frac{i c_2}{4t+c_1} x  \right] \nonumber \\
& \times & \exp \left[ i \left( \frac{n \pi}{L} \right)^2 \frac{1}{4(4t +c_1)}  +  \frac{i~c_2^2}{4 (4t +c_1)} \right].
\end{eqnarray}   

\begin{figure}[t]
\begin{center}
  \includegraphics[width=.35\textwidth]{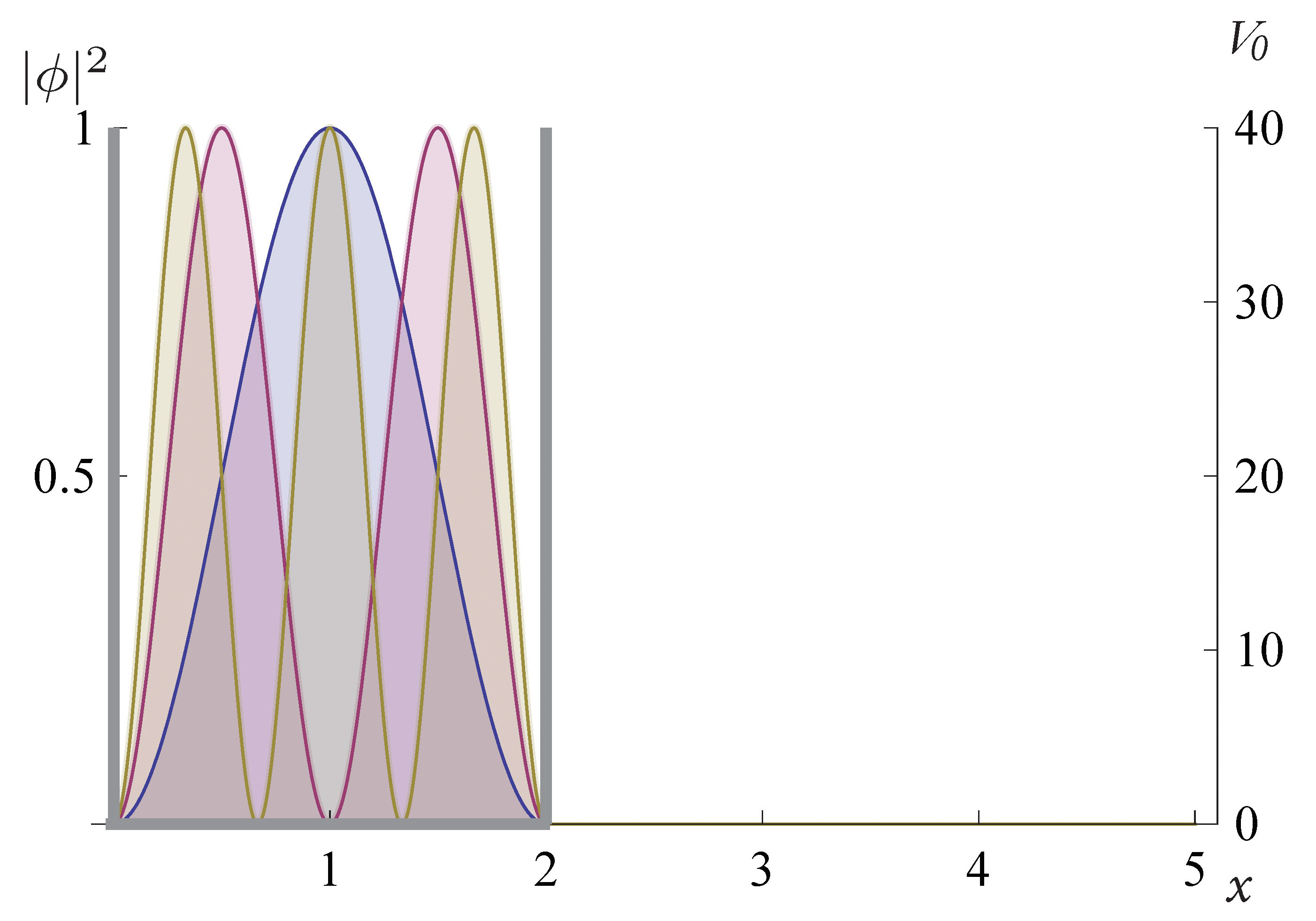} \qquad
  \includegraphics[width=.35\textwidth]{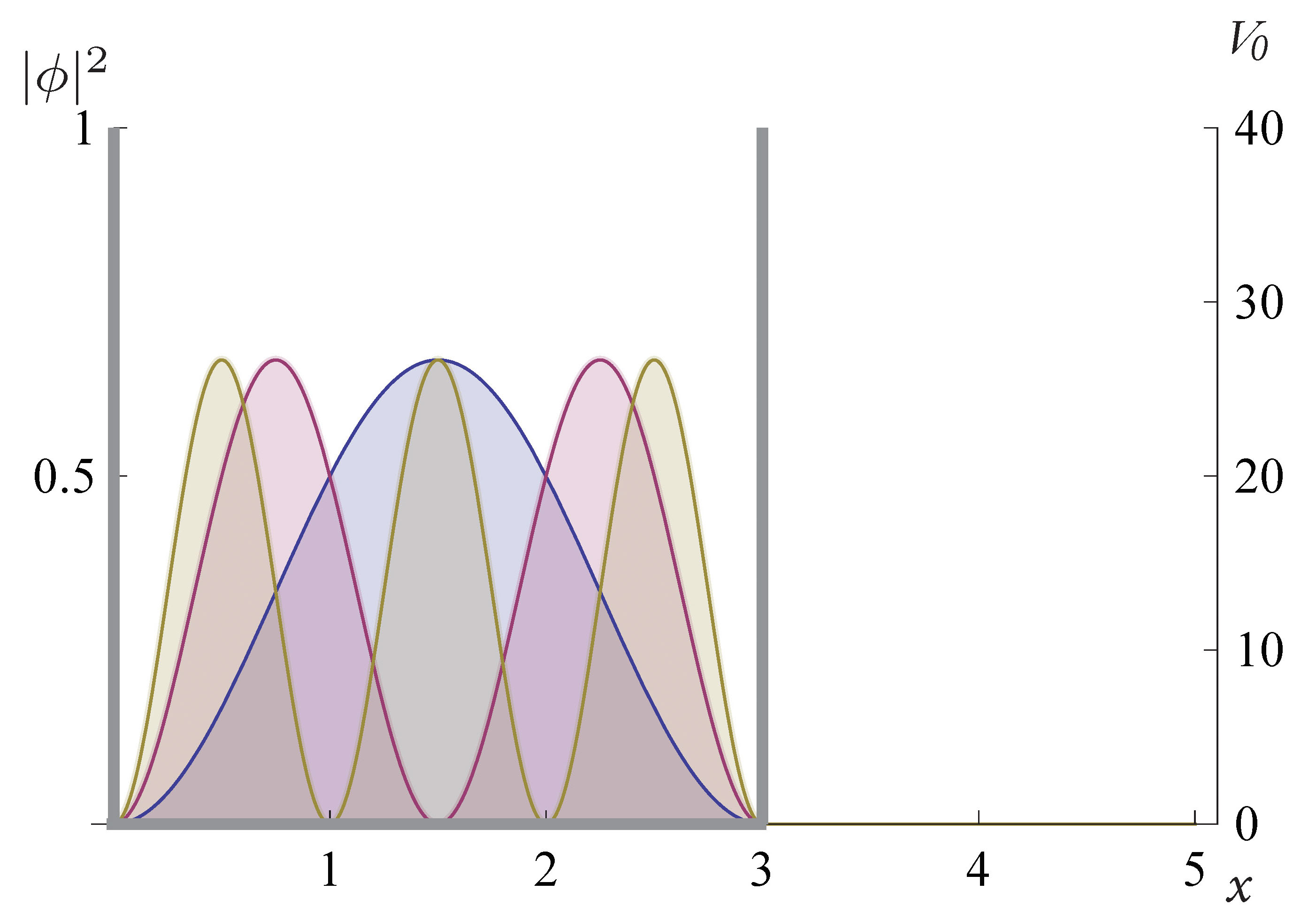}\\
  \includegraphics[width=.35\textwidth]{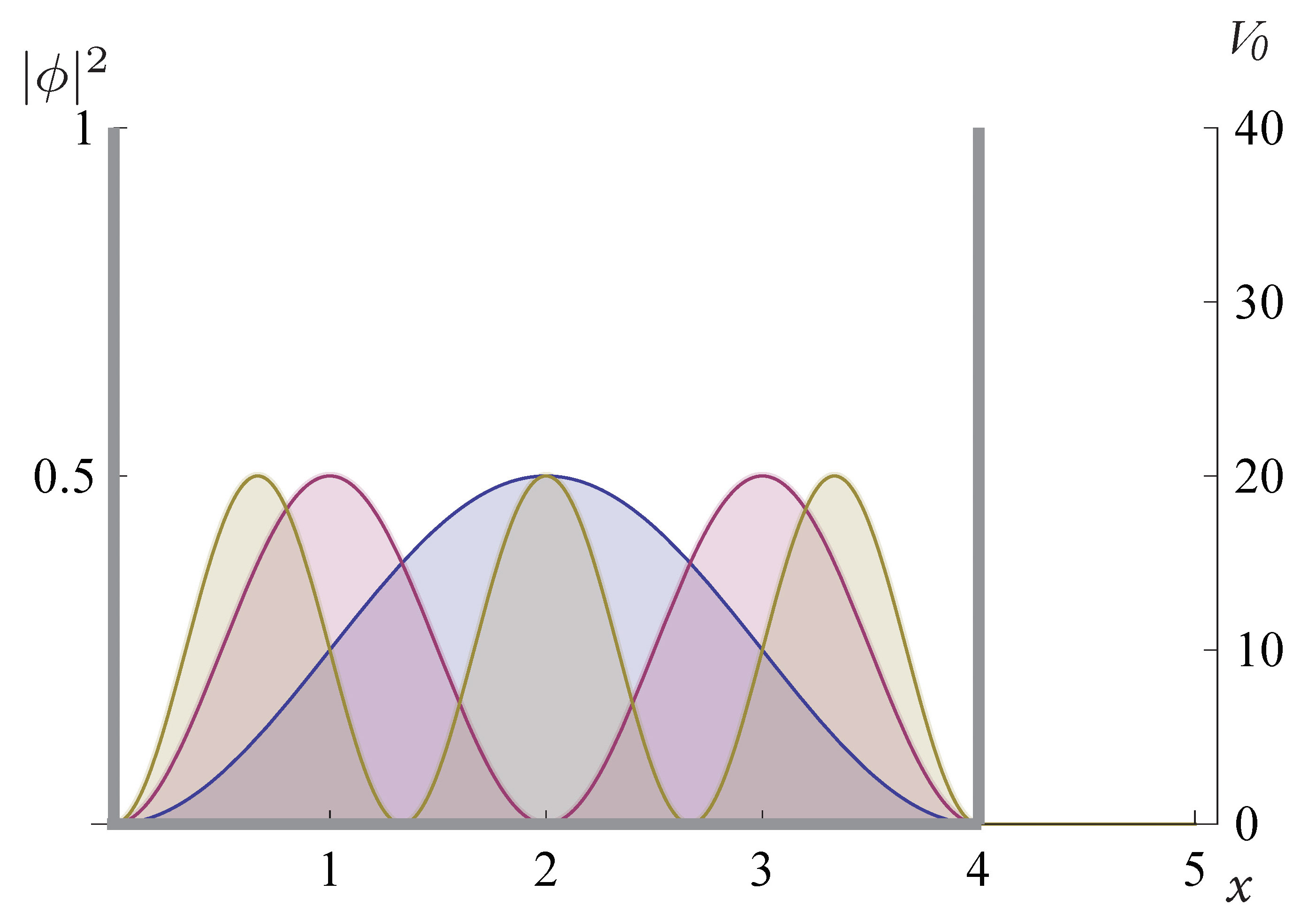} \qquad
  \includegraphics[width=.35\textwidth]{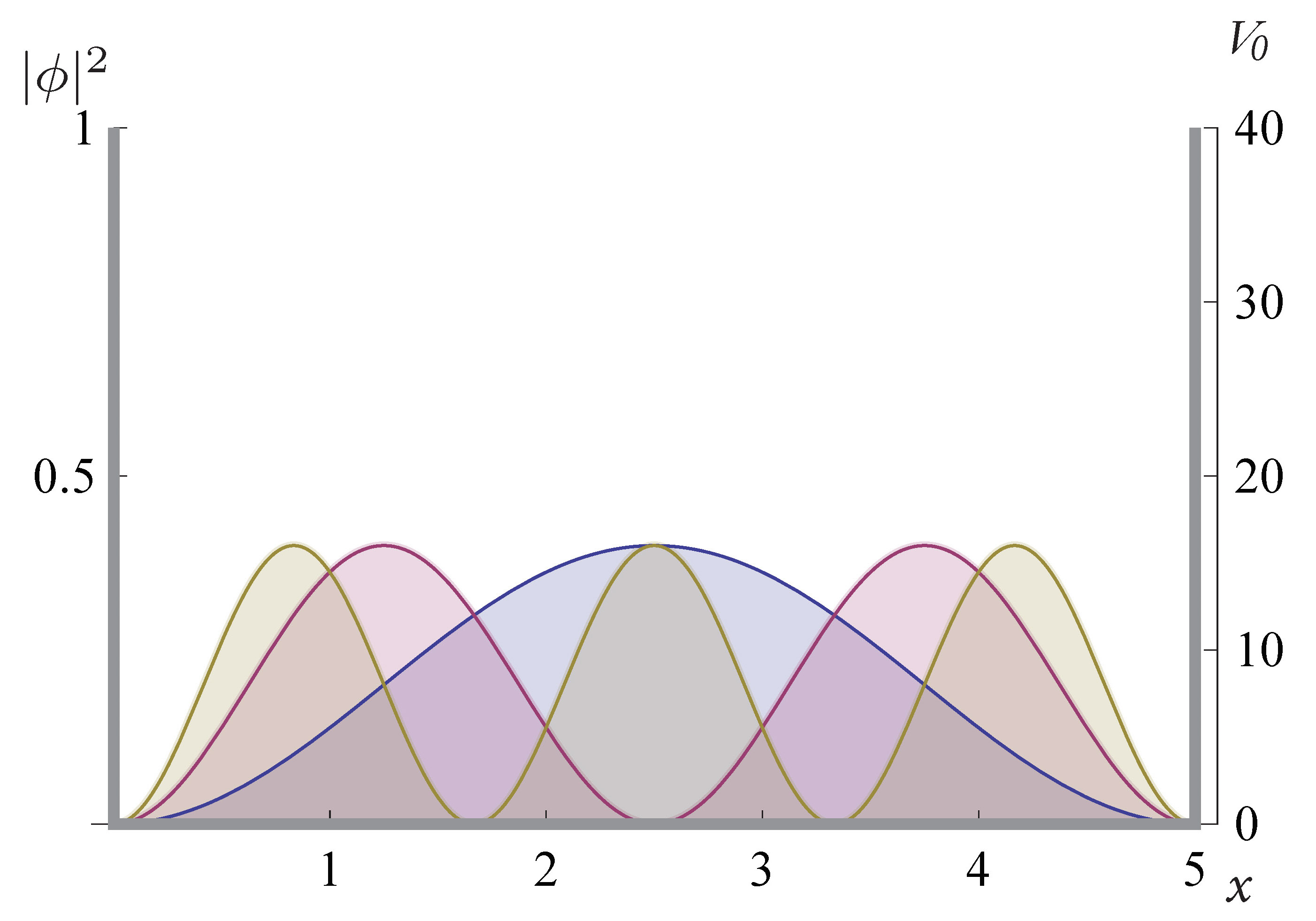}
\caption{Infinite square-well potential with a moving barrier, see \eqref{square moving simplified}. Plot of the probability densities, see \eqref{square moving sol simplified},  $|\phi_1|^2$ (blue), $|\phi_2|^2$ (purple), $|\phi_3|^2$ (yellow), at four different times: $t=1/4$ (top left), $t=1/2$ (top right), $t=3/4$ (bottom left), $t=1$ (bottom right), for the parameter $L=1$.}
\label{fig:1} 
\end{center}      
\end{figure}

We will fix the constant $c_2=0$ so that one barrier is always at zero. We take as well $c_1=1$, then the moving barrier will be at $L$ when $t=0$. The singularity of the problem will be located at $t=-1/4$. For this selection of constant the functions $A$, $B$ in \eqref{ABC} and the change of variable $y$ in \eqref{yvalue} simplify to 
\begin{eqnarray}
A(t)= - \frac{1}{4t+1}, \quad B(t)= 0, \quad y(x,t)= \frac{x}{4 t +1};
\end{eqnarray}
the potential reads 
\begin{align}
V_0(x,t) = \left\{ \begin{array}{lll} 
                0, & \quad & 0 < x < \ell(t), \quad \text{where} \quad \ell(t)= L (4t+1)  \\
                \infty, & \quad & \text{otherwise},
                \end{array} \right. \label{square moving simplified}
\end{align} 
and the solutions of the time dependent Schr\"odinger equation can be written as
\begin{eqnarray}
\phi_n(x,t)= \sqrt{\frac{2}{\ell}} \sin \left( \frac{n \pi}{\ell} x \right) \exp \left\{ i \frac{L}{\ell}\left[ x^2 + \left( \frac{n \pi}{2 L} \right)^2   \right]  \right\}. \label{square moving sol simplified}
\end{eqnarray}  
Note that $\phi_n(0,t)=\phi_n(\ell,t)=0$, satisfying the required boundary conditions of the physical problem. For this specific selection of the constants $c_1$ and $c_2$ the domain of the time variable for the contracting box is $(-\infty,-1/4)$ and for the expanding well is $(-1/4,\infty)$.  Functions (\ref{square moving sol simplified}) are normalized $\int_0^\ell |\phi_n |^2 dx=1$ at any given time. They form a complete orthogonal set at any fixed time and the expectation value of the energy $\langle E \rangle_{\phi_n}(t) = \int_0^\ell \phi_n^* (-\partial_x^2 \phi_n) dx = (n \pi/\ell)^2$.  System (\ref{square moving simplified}) and its solutions (\ref{square moving sol simplified}) were also discussed in \cite{Doescher69,Pinder,Glasser,Gadella}. In Fig. \ref{fig:1} three different probabilities densities were plotted at four different times, see \eqref{square moving sol simplified}: in blue $|\phi_1|^2$, in purple $|\phi_2|^2$ and in yellow $|\phi_3|^2$; for the times $t=1/4, \ 1/2, \ 3/4, \ 1$ and the parameter $L=1$. Since the times we used are greater than $t_0=-1/4$, the plotted potential represents an expanding box.

%

\section{From the infinite well potential with a moving barrier to a P\"oschl-Teller potential with a moving barrier} \label{Sec 3}

In this section we introduce our second tool, a time dependent SUSY transformation introduced in \cite{Matveev,Bagrov95}, the notation is adopted from \cite{contreras17}. Then, we apply it to the infinite well potential with a moving barrier to generate a P\"oschl-Teller potential with a moving barrier.

\subsection{Time dependent supersymmetric quantum mechanics}

We start out with a time dependent Schr\"odinger equation \eqref{Time SE} where the potential $V_0$ is a real known function. Next, we propose the existence of an operator $\mathcal{L}_1$ intertwining two Schr\"odinger operators 
\begin{eqnarray}
S_1 \mathcal{L}_1 = \mathcal{L}_1 S_0,   \label{intertwining 1}
\end{eqnarray}
where the Schr\"odinger operators are defined as $S_j = i \partial_t + \partial_{x}^2 - V_j$, $j=0,1$. Now, if $\mathcal{L}_1$ is a differential operator of the form $\mathcal{L}_1=  A_1 \left(-\partial_x + \frac{u_x}{u} \right)$ where $A_1=A_1(t)$, $u=u(x,t)$ and the subindex in $u_x$ represents partial derivation with respect to $x$, then the intertwining relationship (\ref{intertwining 1}) and the form of the Schr\"odinger operators impose the conditions: 
\begin{eqnarray}
V_1 = V_0 + i (\ln A_1)_t - 2 ( \ln u)_{xx},  \label{V1a} \qquad 
i \partial_t u + u_{xx}-V_0 u = c(t), 
\end{eqnarray}
where $c(t)$ is an integration function. This function $c(t)$ can be absorbed in the potential term, and it will be reflected in the solution $u$ of the Schr\"odinger equation as a time dependent phase, in this work it will be set $c(t)=0$. Note that now $u$ satisfies $S_0 u = 0$, i.e. it is a solution of the initial system. Furthermore, it can be seen from (\ref{V1a}) that in order to avoid new singularities in $V_1$, the functions $A_1$ and $u$ must not vanish.

The potential $V_1$ in \eqref{V1a} is in general a complex function. Since we are interested in a Hermitian operator $S_1$, we must ask that the imaginary part of $V_1$ vanishes, Im$(V_1)=0$. Taking (\ref{V1a}) and considering $V_0$ as a real function, $A_1$ and $u$ must also satisfy $ i \left(\ln |A_1|^2 \right)_t = 2\left(\ln(u/u^*)\right)_{xx}$, since the left hand side depends only on time we can say that     
\begin{eqnarray}
\frac{\partial^3}{\partial x^3} \ln \left( \frac{u}{u^*} \right) = 0 \label{real con}
\end{eqnarray}
is a reality condition to generate a Hermitian operator $S_1$, and then $|A_1|$ is fixed to 
\begin{eqnarray}
|A_1| =  \exp \left\{2 \int \text{Im}\left[\frac{\partial^2}{\partial x^2} \ln u(x,t) \right]dt \right\}. \label{A1}
\end{eqnarray}
If this condition is inserted into (\ref{V1a}) along with $A_1=|A_1|$, then the expression of the new potential simplifies to  
\begin{eqnarray}
V_1=V_0 -  \frac{\partial^2}{\partial x^2} \ln |u|^2. \label{V1}
\end{eqnarray} 

The intertwining relation (\ref{intertwining 1}) ensures that if $\phi$ solves the equation $S_0 \phi = 0$, then $\chi = \mathcal{L}_1 \phi$ solves $S_1 \chi = 0$. Direct substitution shows that there is an extra function $\chi_\epsilon=1/A_1 u^*$, often called missing state, that also solves $S_1 \chi_\epsilon=0$. Consult \cite{Matveev,Bagrov95,contreras17,Zelaya17} for more details on this technique.

\subsection{Trigonometric P\"oschl-Teller potential with a moving barrier}
\begin{figure}[t]
\begin{center}
  \includegraphics[width=.35\textwidth]{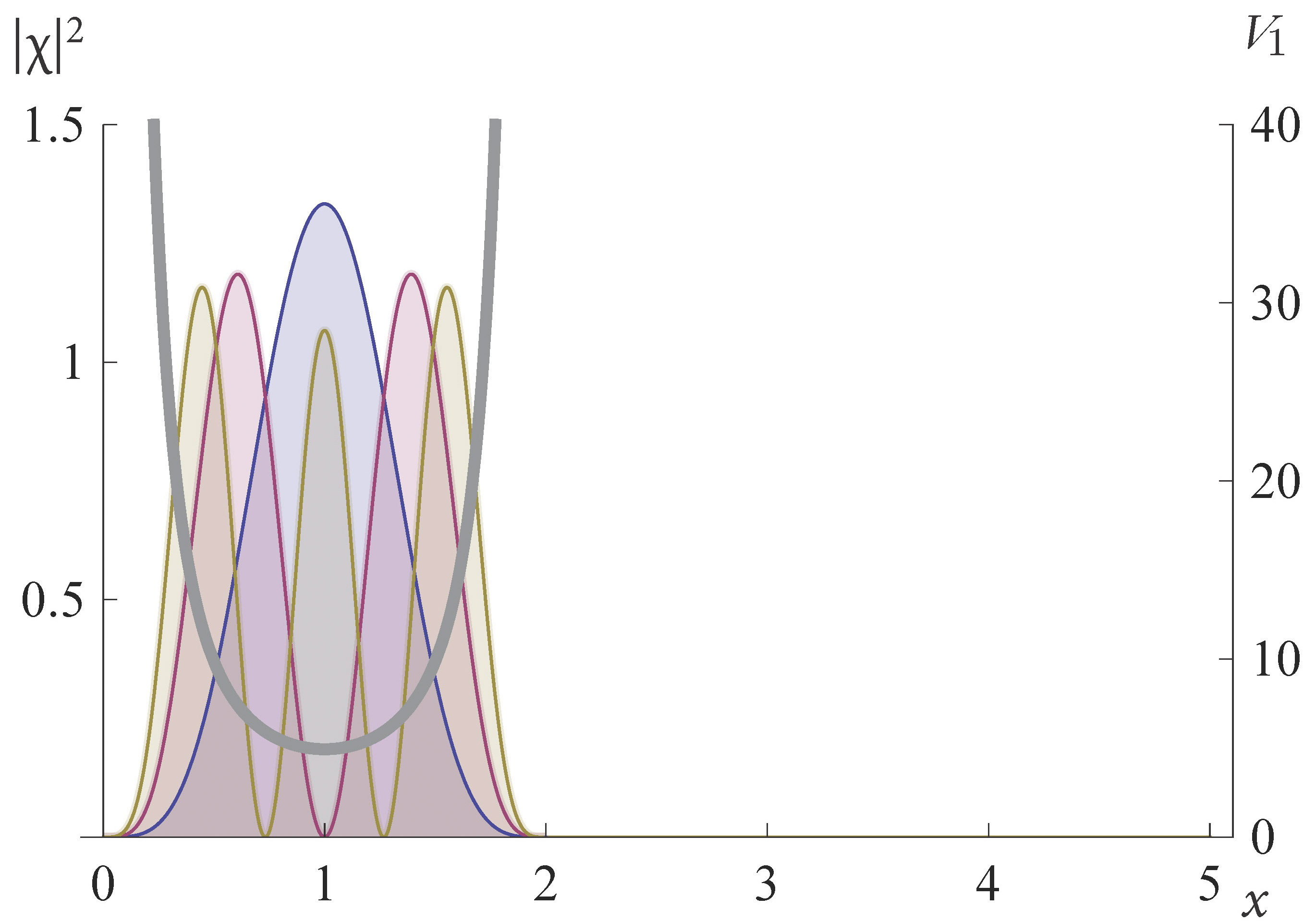} \qquad
  \includegraphics[width=.35\textwidth]{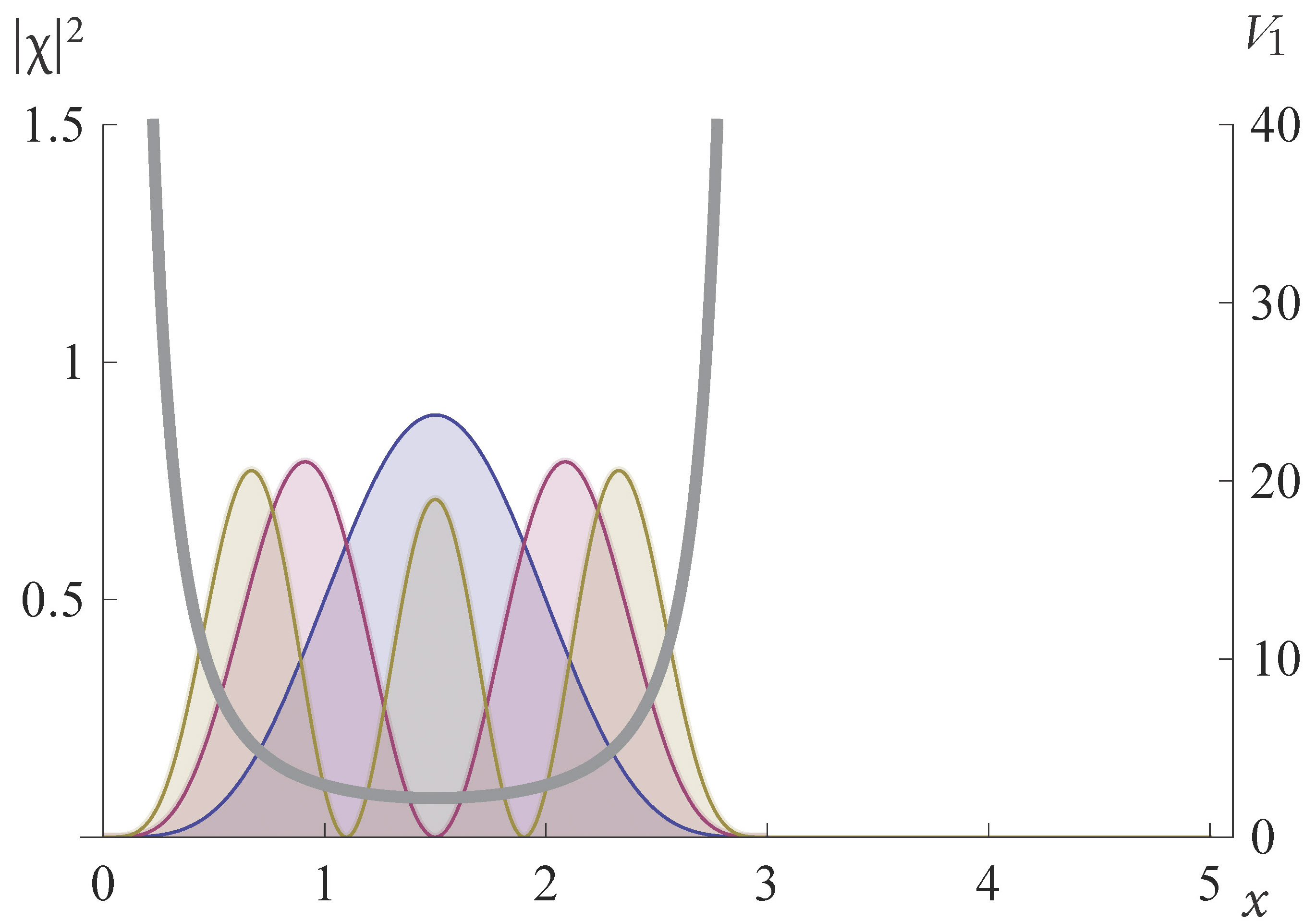} \\
  \includegraphics[width=.35\textwidth]{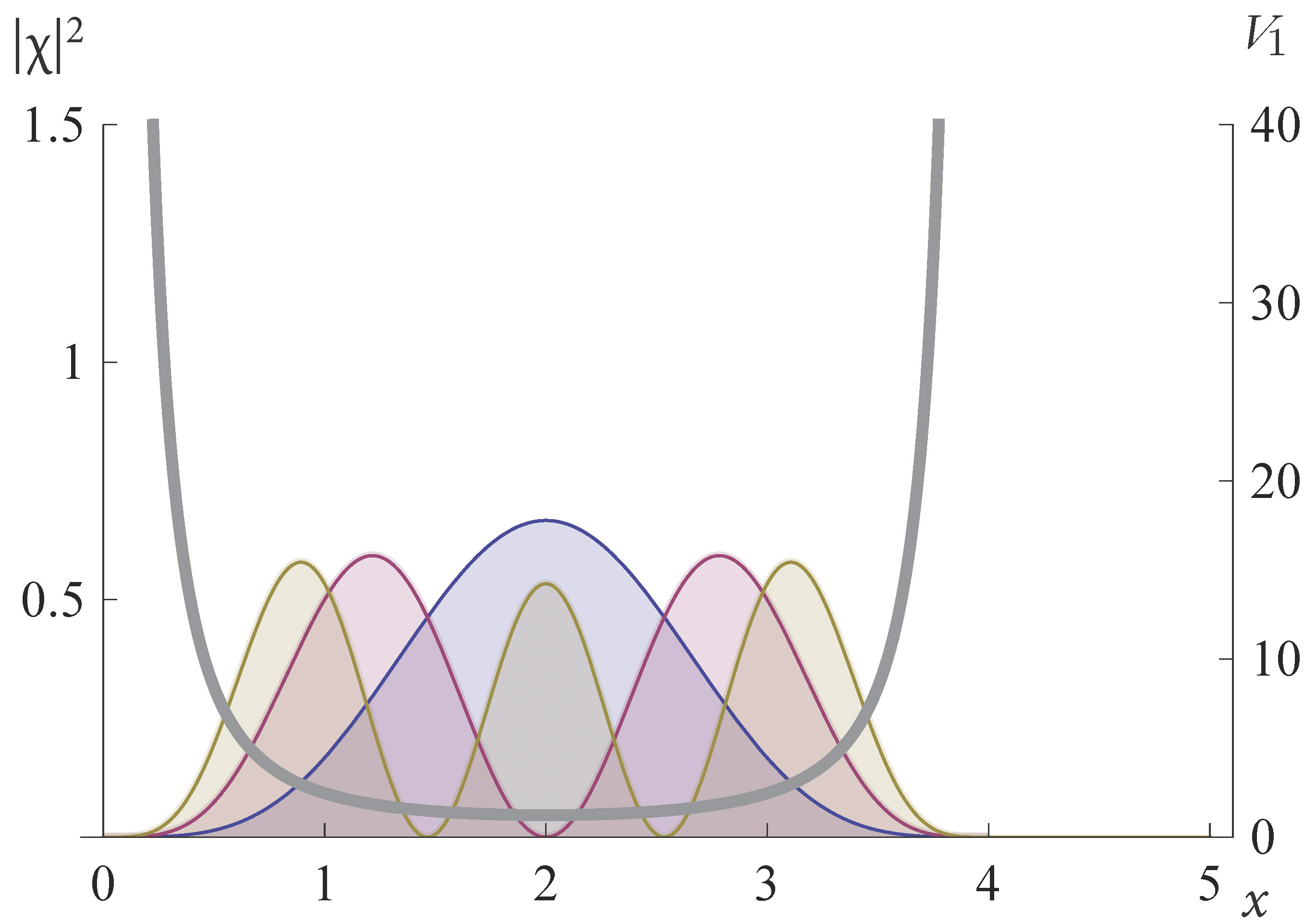} \qquad
  \includegraphics[width=.35\textwidth]{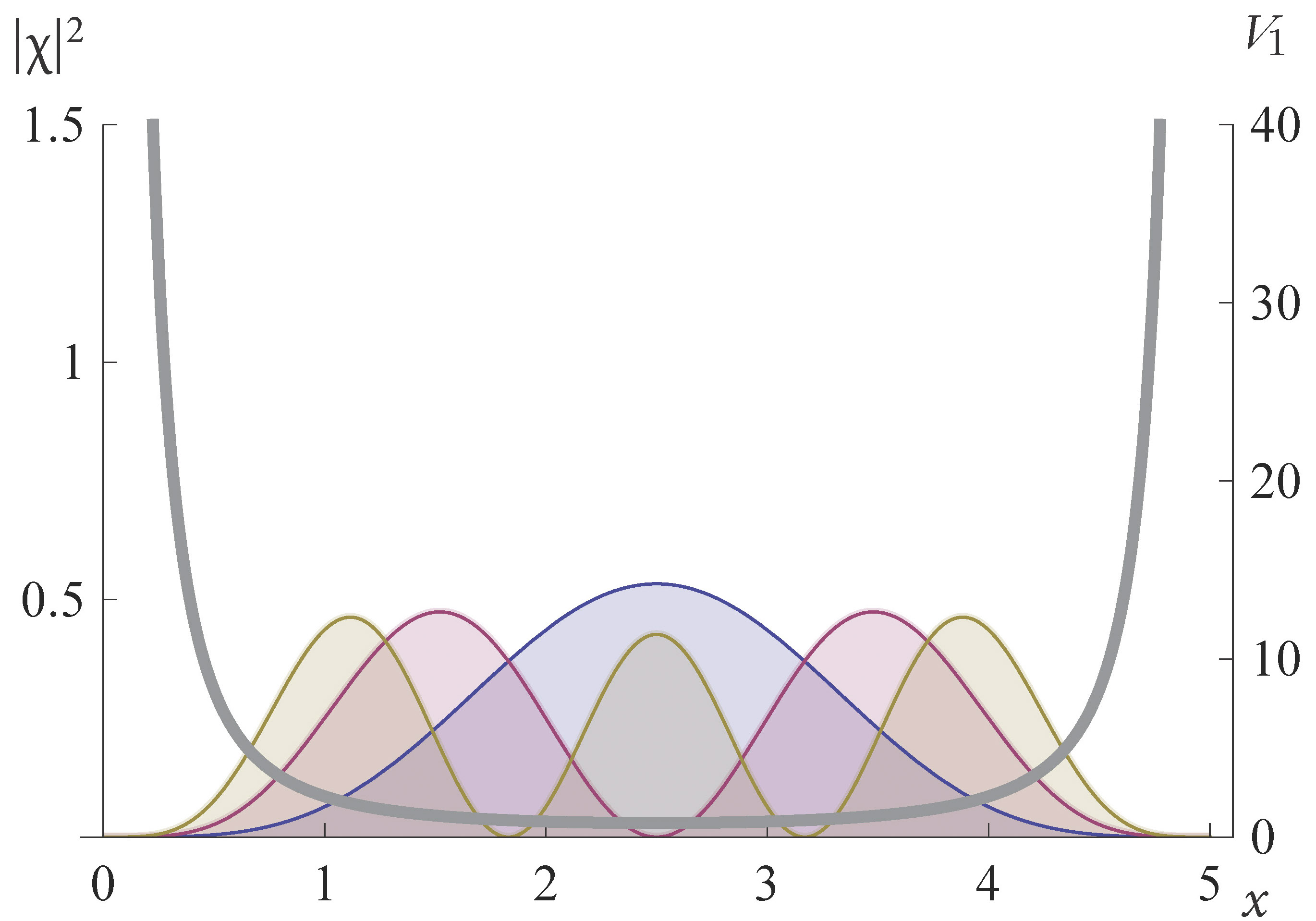}
\caption{In gray a trigonometric P\"oschl-Teller potential with a moving barrier, see \eqref{poschl-teller moving}. Moreover, normalized probability densities are also plotted, see \eqref{chi moving},  $|\chi_2|^2$ (blue), $|\chi_3|^2$ (purple), $|\chi_4|^2$ (yellow), at four different times: $t=1/4$ (top left), $t=1/2$ (top right), $t=3/4$ (bottom left), $t=1$ (bottom right), for the parameter $L=1$.}
\label{fig:2}
\end{center}     
\end{figure}

In order to apply a 1-SUSY transformation to the time dependent potential defined in \eqref{square moving simplified}, we need to select a transformation function $u(x,t)$ fulfilling three conditions: \textit{i)} $u(x,t)$ must satisfy the time dependent Schr\"odinger equation $S_0 u=0$, \textit{ii)} $u(x,t) \neq 0$  to avoid new singularities inside the domain of the potential and \textit{iii)} $\partial_x^3 \ln( u/u^*)=0$ to generate a Hermitian potential $V_1$. One function satisfying all three conditions is $\phi_1(x,t)$ in \eqref{square moving sol simplified}, thus, we will use it as transformation function:  
\begin{eqnarray}
u(x,t)= \sqrt{\frac{2}{\ell}} \sin \left( \frac{ \pi}{\ell} x \right) \exp \left\{ i \frac{L}{\ell}\left[ x^2 + \left( \frac{\pi}{2 L} \right)^2   \right]  \right\} , \qquad \ell=L(4t+1).
\end{eqnarray}  
Then, we need to calculate the function $A_1$, see (\ref{A1}), and the intertwining operator $\mathcal{L}_1=  A_1 \left(-\partial_x + \frac{u_x}{u} \right)$:    
\begin{eqnarray}
A_1(t)= 4t+1, \qquad \mathcal{L}_1=- (4t+1) \frac{\partial}{\partial_x} +i 2 x +  \frac{\pi}{L}\cot \left(\frac{\pi x}{\ell} \right).  \label{Poschl operator}  
\end{eqnarray}
The 1-SUSY partner $V_1$ of \eqref{square moving simplified} can be obtained directly from (\ref{V1}) as
\begin{align}
V_1(x,t) = \left\{ \begin{array}{lll} 
                2 \left( \frac{\pi}{\ell} \right)^2 \csc^2\left( \frac{\pi x}{\ell} \right), & \quad & 0 < x < \ell(t),    \\
                \infty, & \quad & \text{otherwise},
                \end{array} \right. \label{poschl-teller moving}
\end{align}
it coincides with a trigonometric P\"oschl-Teller potential at any fixed time \cite{Contreras08}, emphasizing that in our situation the potential has a moving wall. Solutions of the time dependent Schr\"odinger equation for this potential can be obtain applying the operator $\mathcal{L}_1$ onto solutions $\phi_n$, see \eqref{square moving sol simplified} and \eqref{Poschl operator}: 
\begin{align}
& \chi_n(x,t)= \mathcal{L}_1 \phi_n(x,t) \nonumber \\ 
&= \frac{\pi}{L}\sqrt{\frac{2 }{\ell}}  \left[\cot\left(\frac{\pi x}{\ell} \right) \sin\left( \frac{n \pi x}{\ell} \right) -n \cos\left( \frac{n \pi x}{\ell} \right)   \right] \exp \left\{ i \frac{L}{\ell}\left[ x^2 + \left( \frac{n \pi}{2 L} \right)^2   \right]  \right\}, \label{chi moving}
\end{align}
where $n=2,3,4,\cdots$. In this problem $\lim_{x\to 0} \chi_n(x,t)=\lim_{x\to \ell} \chi_n(x,t)=0$. There is no square integrable missing state $\chi_\epsilon$. In Fig. \ref{fig:2} the P\"oschl-Teller potential with a moving barrier and the normalized probability densities corresponding to $\chi_2$, $\chi_3$ and $\chi_4$ are shown at four different times $t=1/4, ~1/2,~3/4,~1$, for the parameter $L=1$.

\section{Confluent SUSY partners: more potentials with a moving barrier} \label{Sec 4}

The 1-SUSY technique introduced in Sec. \ref{Sec 3} has as constraint that the transformation function $u$ must never vanish. To underpass this restriction a second iteration can be performed, the particular iteration we will use is known as confluent SUSY, see \cite{contreras17} for the time dependent version and \cite{Fernandez03} for the time independent case. This technique will be applied again to $V_0$ and will generate a new family of infinite potential wells with a moving barrier. 

\subsection{Time dependent confluent SUSY}

Departing from $S_1$ we propose a second intertwining operator $\mathcal{L}_2$ connecting $S_1$ with a new Schr\"odinger operator $S_2$ 
\begin{eqnarray}
S_2\mathcal{L}_2 = \mathcal{L}_2 S_1, \label{intertwining 2}
\end{eqnarray}
where again $\mathcal{L}_2$ is a differential operator on the form $\mathcal{L}_2 = A_2\left(-\partial_x+\frac{v_x}{v}  \right)$, where $v$ solves $S_1 v =0$. We would like to use a function $v$ written in term of $u$. If the missing state $\chi_\epsilon = 1/A_1u^*$ is used, then the generated potential $V_2$ is exactly the initial potential $V_0$. In order to generate a different potential we should use a more general solution $v$:

\begin{eqnarray}
v=\frac{1}{A_1 u^*} \left( \omega +  \int^x_{x_0} |u(s,t)|^2 ds  \right), \label{v 2SUSY}
\end{eqnarray}
where $\omega$ is a real constant. It can be verified by direct substitution that this general expression for $v$ is indeed a solution of $S_1 v =0$. We can also demand $S_2$ and $S_1$ to be Hermitian operators. This directly implies $\partial^3_x \ln(v/v^*)=0$, and substituting (\ref{v 2SUSY}), the Hermiticity condition for the second potential is also $\partial^3_x \ln(u/u^*)=0$.  Analogous to (\ref{A1}),  since $\omega$ was chosen real, $A_2$ can be fixed as $A_2=A_1$. Under these considerations the new potential is given by
\begin{eqnarray}
V_2 = V_1 - 2\partial_{xx}\ln |v| = V_0 - 2\partial_{xx} \ln \left( \omega + \int^x_{x_0} |u(s,t)|^2 ds \right). \label{new pot}
\end{eqnarray}
From the intertwining relation (\ref{intertwining 2}) and using the functions $\chi_n$ (solving $S_1 \chi=0$), we can see that functions  $\xi_n = \mathcal{L}_2 \chi_n= \mathcal{L}_2 \mathcal{L}_1 \phi_n$ will solve the equation $S_2 \xi =0$. Finally, a missing solution can be found as 
\begin{eqnarray}
\xi_\epsilon=\frac{1}{A_2 v^*}= \frac{u}{ \omega +  \int^x_{x_0} |u(s,t)|^2 ds  }.  \label{missing confluent}
\end{eqnarray}
 
Confluent and 1-SUSY techniques present similarities. Indeed, both use only one transformation function $u$ fulfilling $S_0 u=0$,  both require  $\partial_x^3 \ln(u/u^*)=0$ to generate Hermitian potentials, but the regularity condition is different. In 1-SUSY $u$ must be nodeless, for confluent SUSY the transformation function satisfies a more relaxed condition: $\int^x_{x_0} |u(s,t)|^2 ds  \neq -\omega$, this last condition could be met, for example, by any square integrable solution.

\subsection{More potentials with a moving barrier}
 
Departing from the infinite well potential with a moving barrier in (\ref{square moving simplified}), we can notice that solutions $\phi_n(x,t)$ (see (\ref{square moving sol simplified})), when $n \geq 2$ cannot be used as transformation  function for a 1-SUSY transformation because they have at least one zero in the interval $(0,\ell)$. With the confluent SUSY algorithm presented in this section we can surpass this restriction. 

By selecting $u(x,t)= \phi_m(x,t)$ (see (\ref{square moving sol simplified})), where $m \in \mathbb{N}$ is a fixed number, a confluent SUSY partner of the infinite square-well potential with a moving barrier can be constructed. First we need to find the function $A_2=A_1=|A_1|$, see \eqref{A1}, and the intertwining operators $\mathcal{L}_1= A_1\left(-\partial_x + u_x/u \right)$ and $\mathcal{L}_2=A_1\left(-\partial_x + v_x/v \right)$: 
\begin{align}
&A_2(t) = A_1(t)=4t+1, \nonumber \\  
& \mathcal{L}_1= - (4t+1) \frac{\partial}{\partial_x} +i 2 x +  \frac{m \pi}{L}\cot \left(\frac{m \pi x}{\ell} \right),  \nonumber \\
&\mathcal{L}_2=- (4t+1) \frac{\partial}{\partial_x} +i 2 x -  \frac{m \pi}{L}\cot \left(\frac{m \pi x}{\ell} \right)+ \frac{4 m \pi \ell \sin^2\left(\frac{m \pi x}{\ell} \right)}{2 m \pi L(x + \ell \omega)- L \ell \sin\left(\frac{2 m \pi x}{\ell} \right)},  \label{Poschl operator 2}  
\end{align}
where we fixed $x_0=0$ in the definition of $v$, see \eqref{v 2SUSY}, \eqref{new pot} and \eqref{missing confluent}. Then, using (\ref{new pot}) an expression for the potential $V_2$ can be obtained:
\begin{align}
V_2(x,t) = \left\{ \begin{array}{lll} 
                 \frac{32 \left(\frac{ m \pi}{\ell}\right)^2  \sin\left( \frac{ m \pi x}{\ell} \right)\left[\sin\left( \frac{ m \pi x}{\ell} \right)- \frac{m \pi}{\ell} \cos\left( \frac{ m \pi x}{\ell} \right) (x + \omega \ell)  \right]}{\left[ \sin\left( \frac{2 m \pi x}{\ell} \right) - 2 \frac{m \pi}{\ell}(x + \omega \ell)  \right]^2} , & \quad & 0 < x < \ell(t),  \\
                \infty, & \quad & \text{otherwise}, 
                \end{array} \right. \label{poschl-teller confluente moving}
\end{align}
where $\omega \in (\infty,-1] \cup [0,\infty)$ is a  constant introduced by confluent algorithm. 

\begin{figure}[t]
\begin{center}
  \includegraphics[width=.35\textwidth]{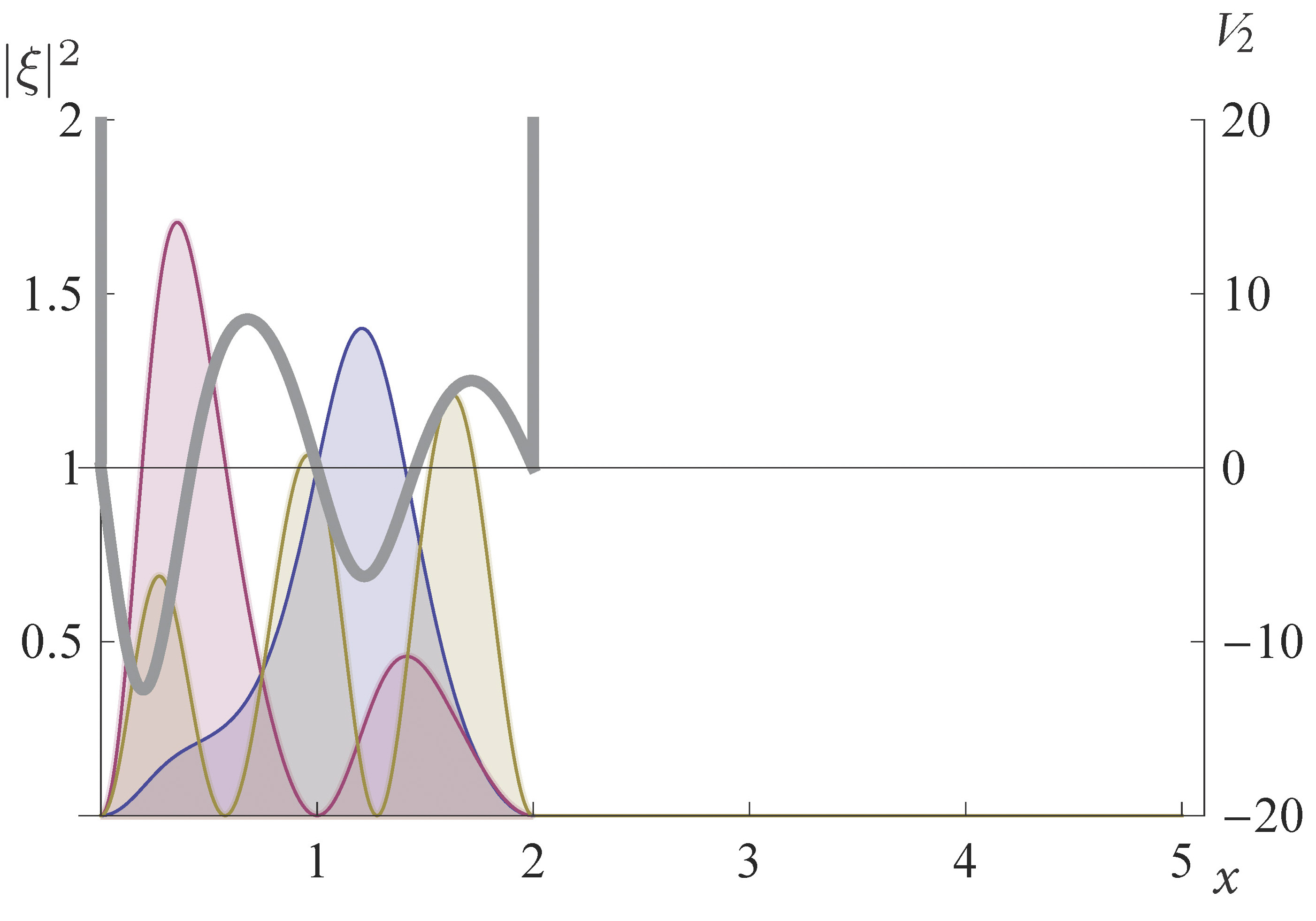} \qquad
  \includegraphics[width=.35\textwidth]{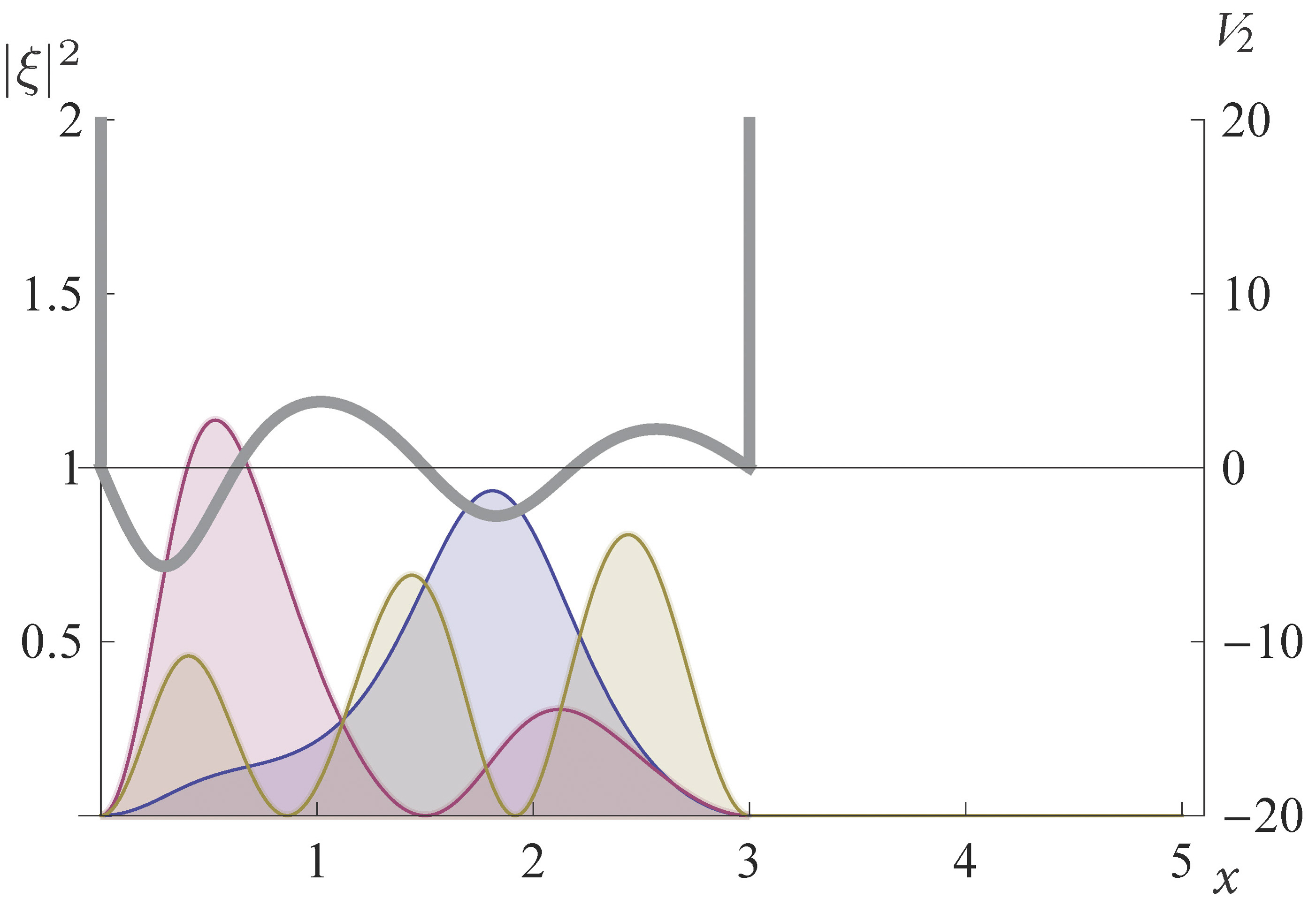} \\
  \includegraphics[width=.35\textwidth]{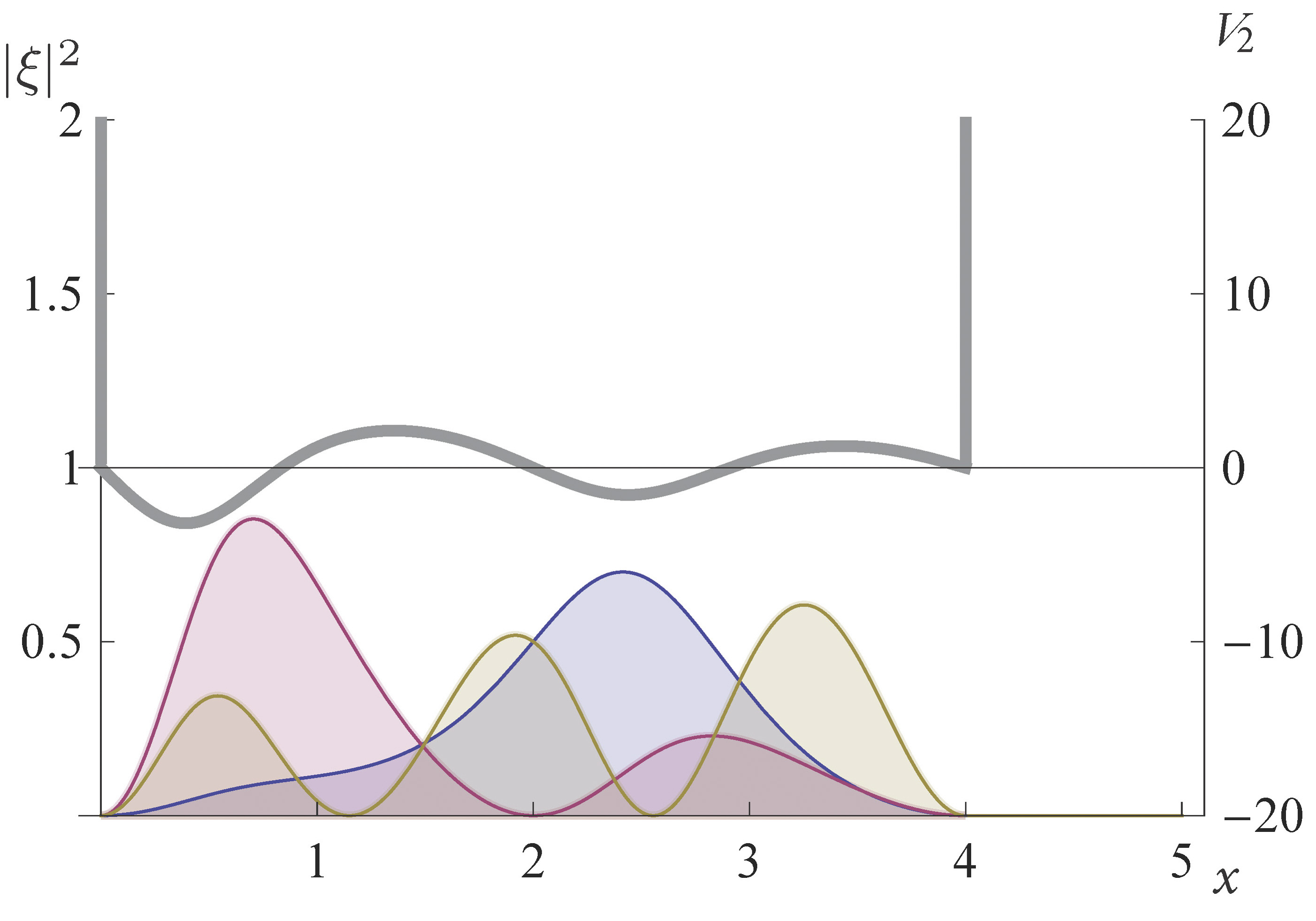} \qquad
  \includegraphics[width=.35\textwidth]{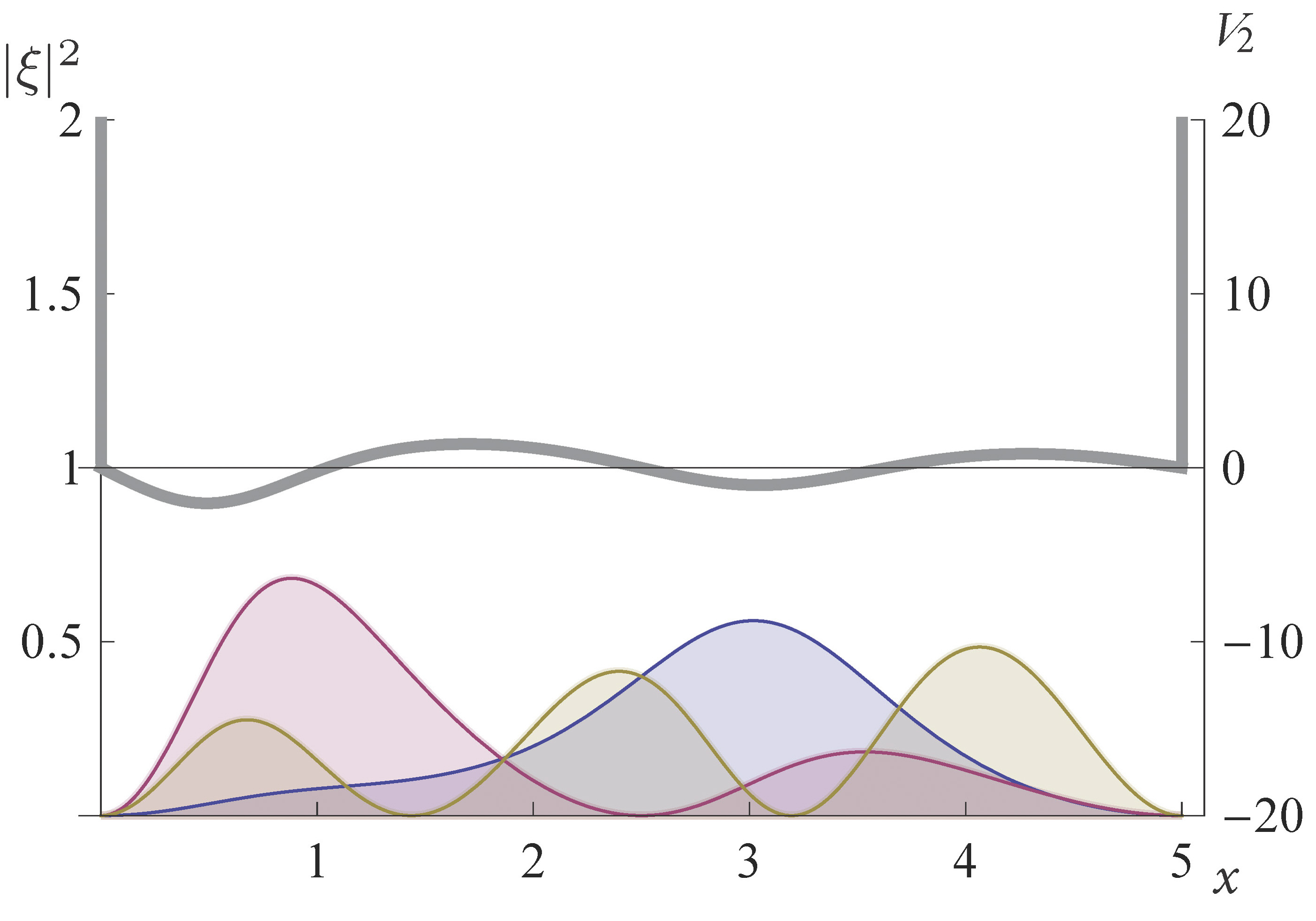}
\caption{In gray a confluent SUSY partner of the infinite square-well potential with a moving barrier, see \eqref{poschl-teller confluente moving}. Moreover,  normalized probability densities are also plotted, see \eqref{xin solutions} and \eqref{xi missing},  $|\xi_1|^2$ (blue), $|\xi_\epsilon|^2$ (purple), $|\xi_3|^2$ (yellow), at four different times: $t=1/4$ (top left), $t=1/2$ (top right), $t=3/4$ (bottom left), $t=1$ (bottom right), for the parameters $L=1$, $\omega=0.4$ and $m=2$.}
\label{fig:3} 
\end{center}      
\end{figure}

\begin{figure}[h]
\begin{center}
  \includegraphics[width=.35\textwidth]{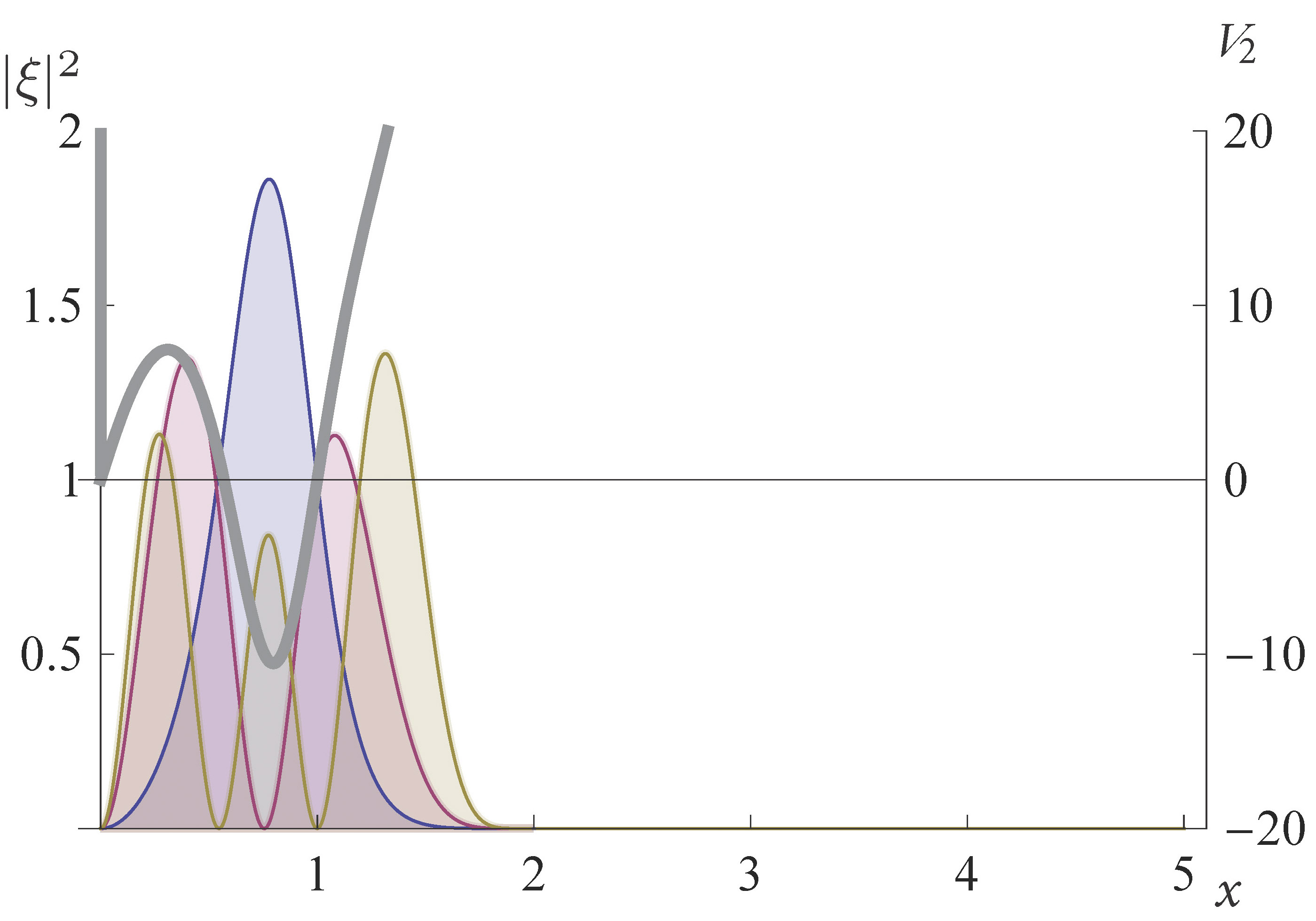} \qquad
  \includegraphics[width=.35\textwidth]{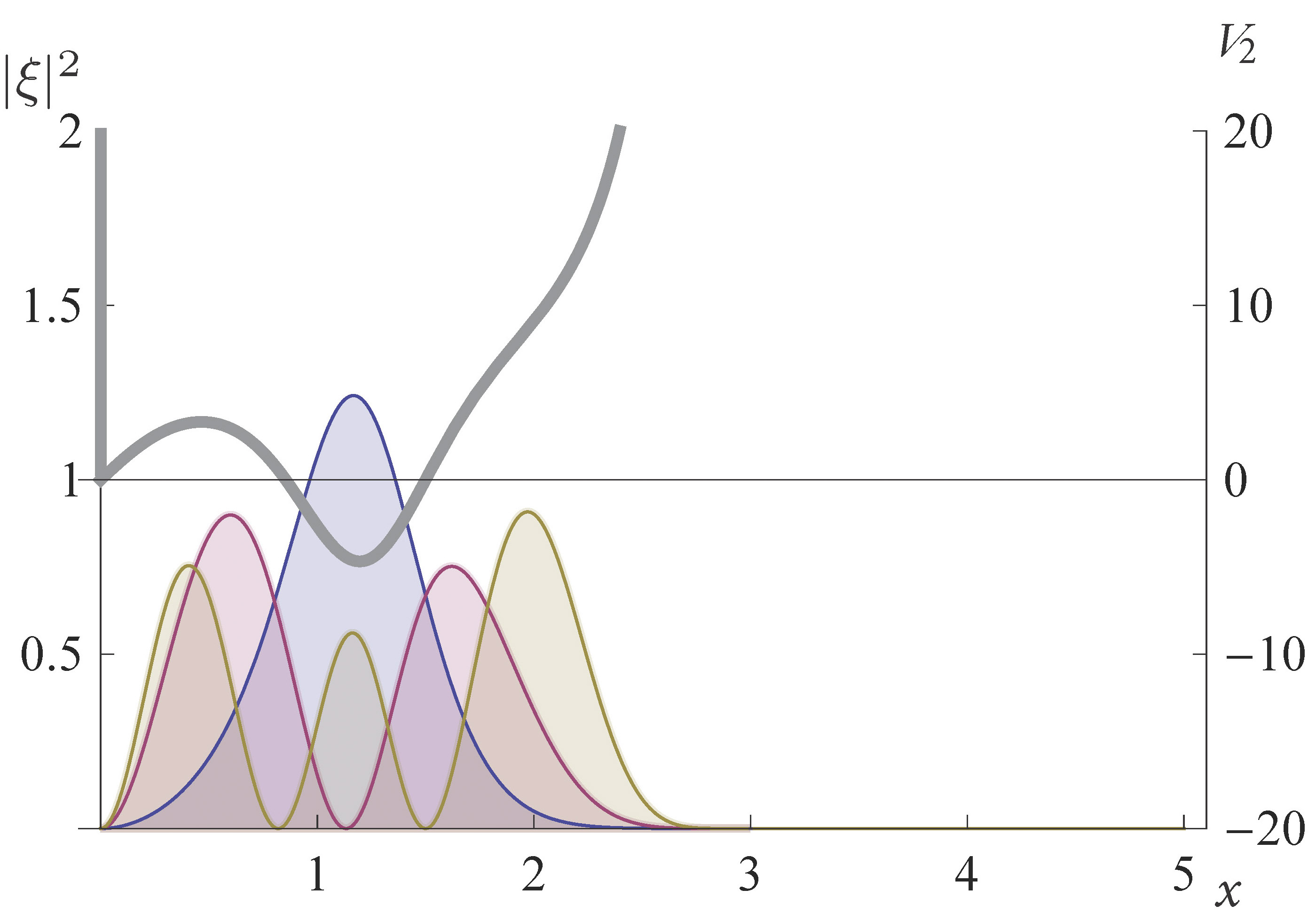} \\
  \includegraphics[width=.35\textwidth]{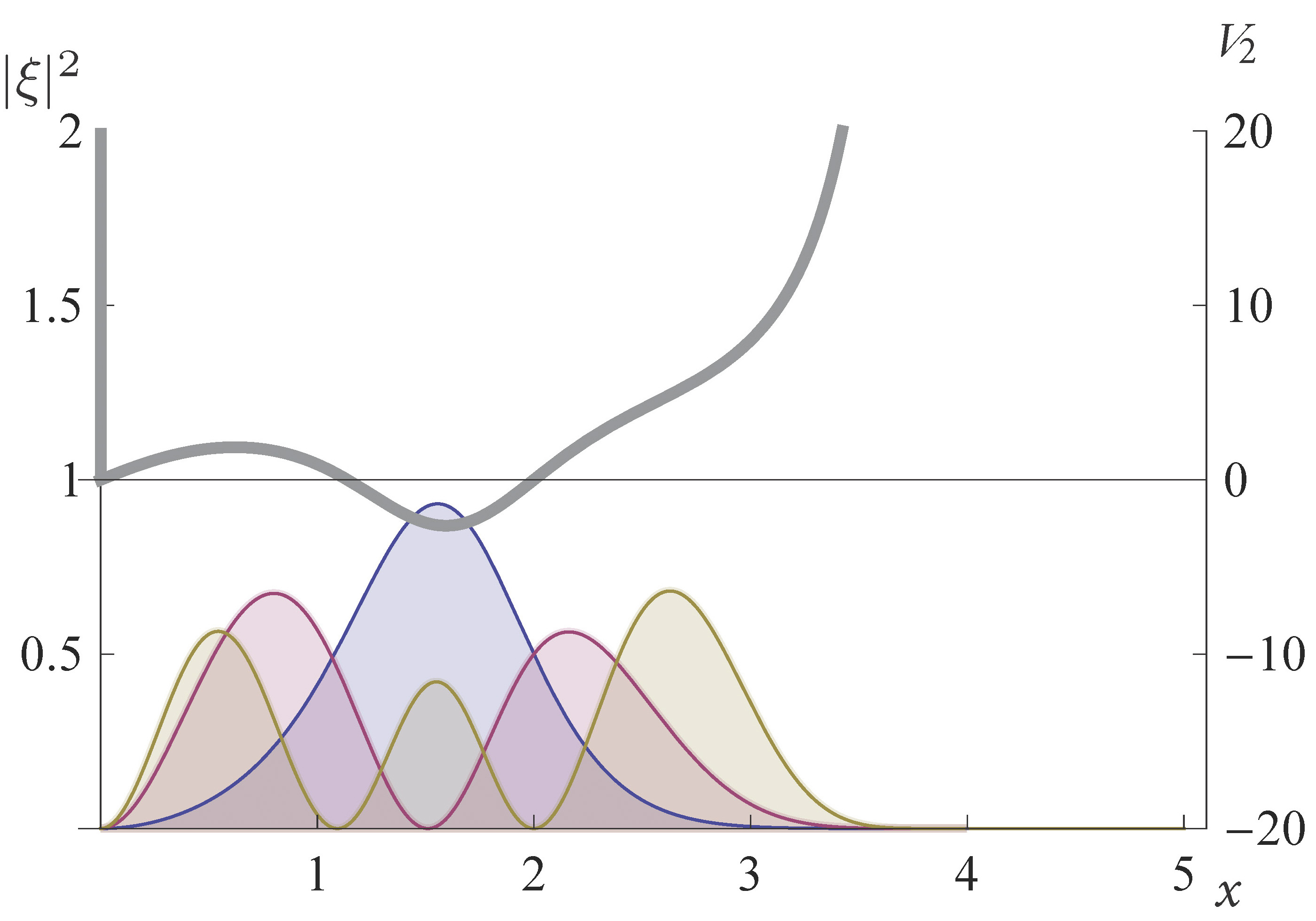} \qquad
  \includegraphics[width=.35\textwidth]{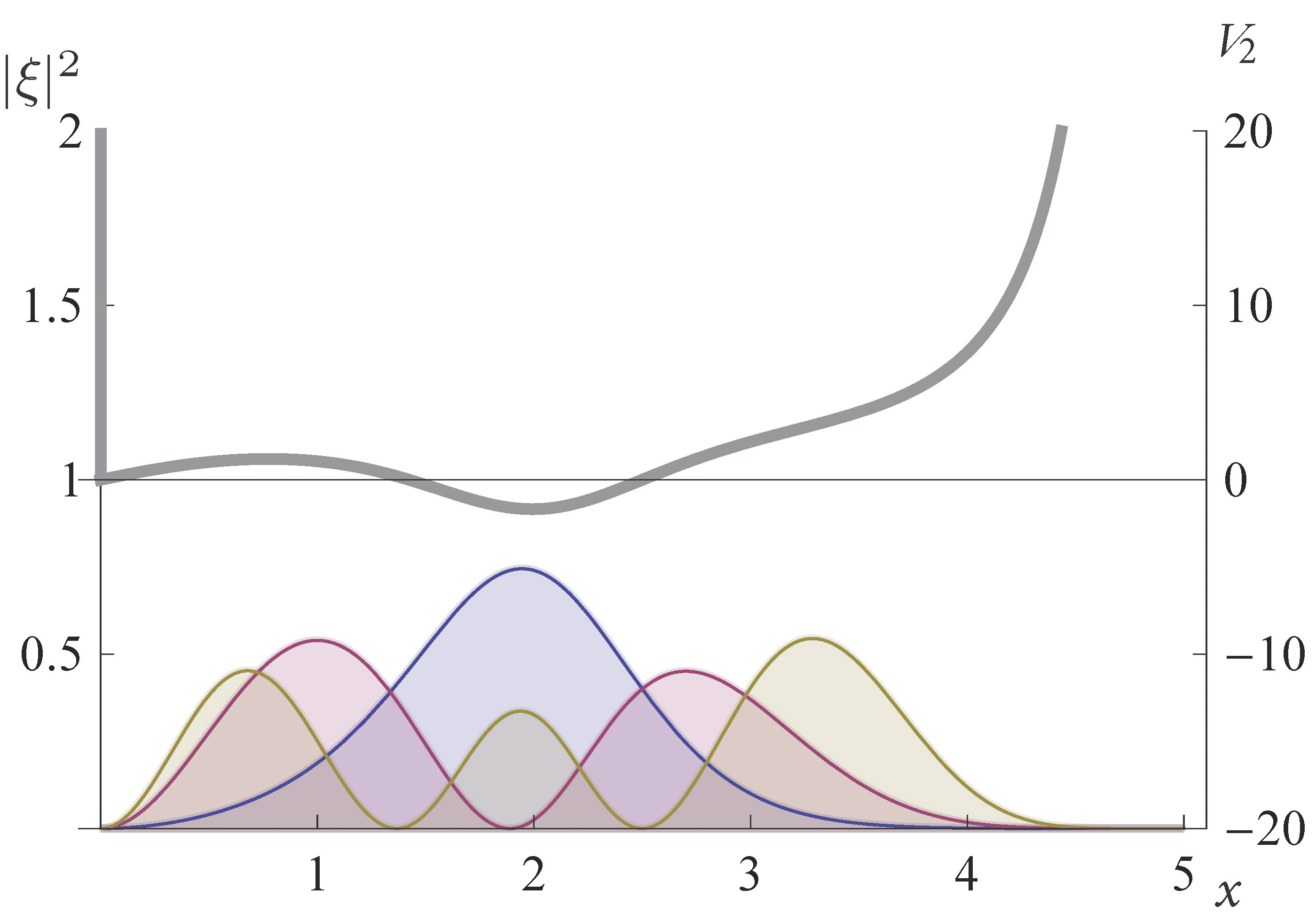}
\caption{In gray a confluent SUSY partner of the infinite square-well potential with a moving barrier, see \eqref{poschl-teller confluente moving}. Moreover,  normalized probability densities are also plotted, see \eqref{xin solutions},  $|\xi_1|^2$ (blue), $|\xi_3|^2$ (purple), $|\xi_4|^2$ (yellow), at four different times: $t=1/4$ (top left), $t=1/2$ (top right), $t=3/4$ (bottom left), $t=1$ (bottom right), for the parameters $L=1$, $\omega=-1$ and $m=2$.}
\label{fig:4} 
\end{center}      
\end{figure}

\begin{figure}[th]
\begin{center}
  \includegraphics[width=.35\textwidth]{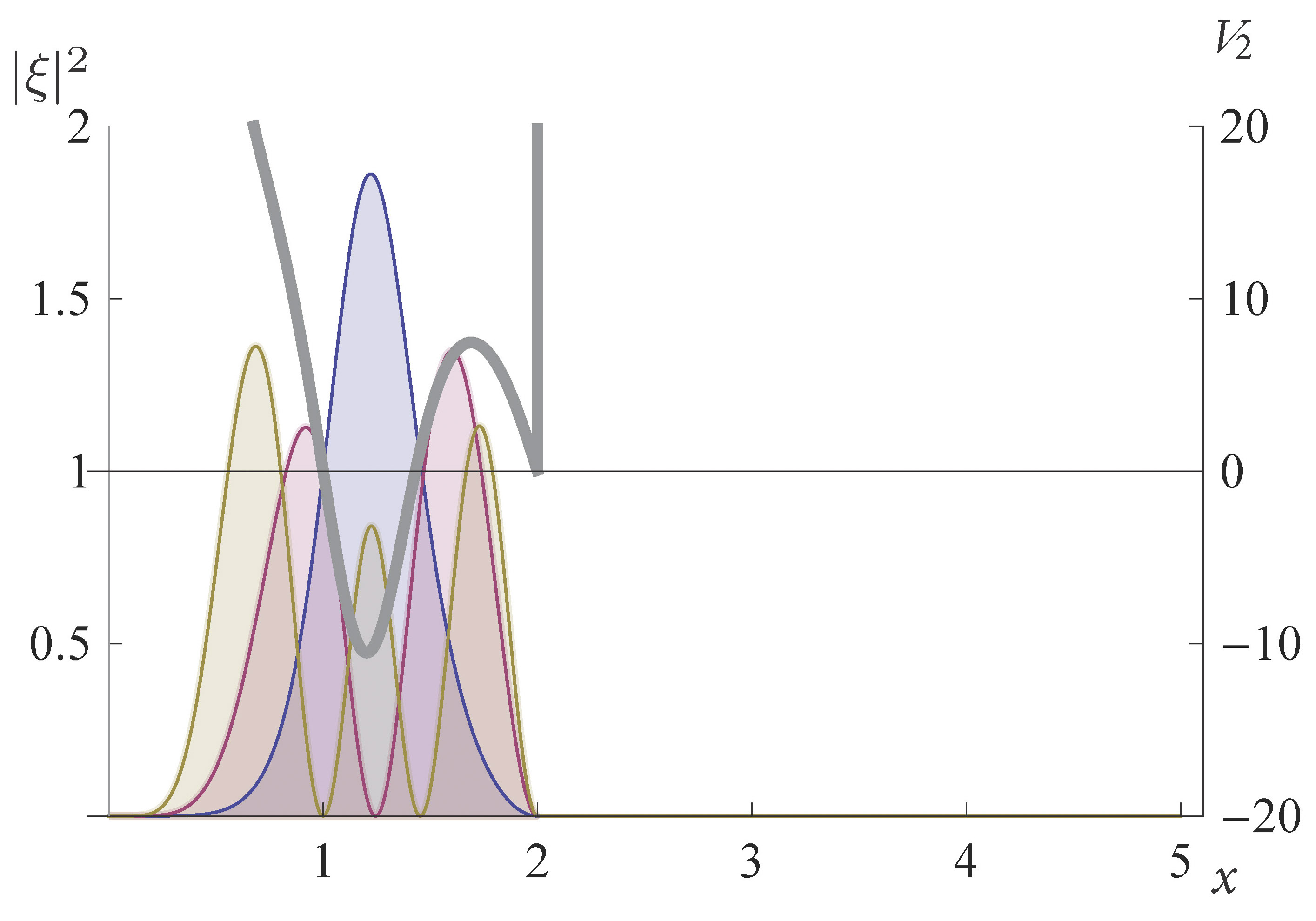} \qquad
  \includegraphics[width=.35\textwidth]{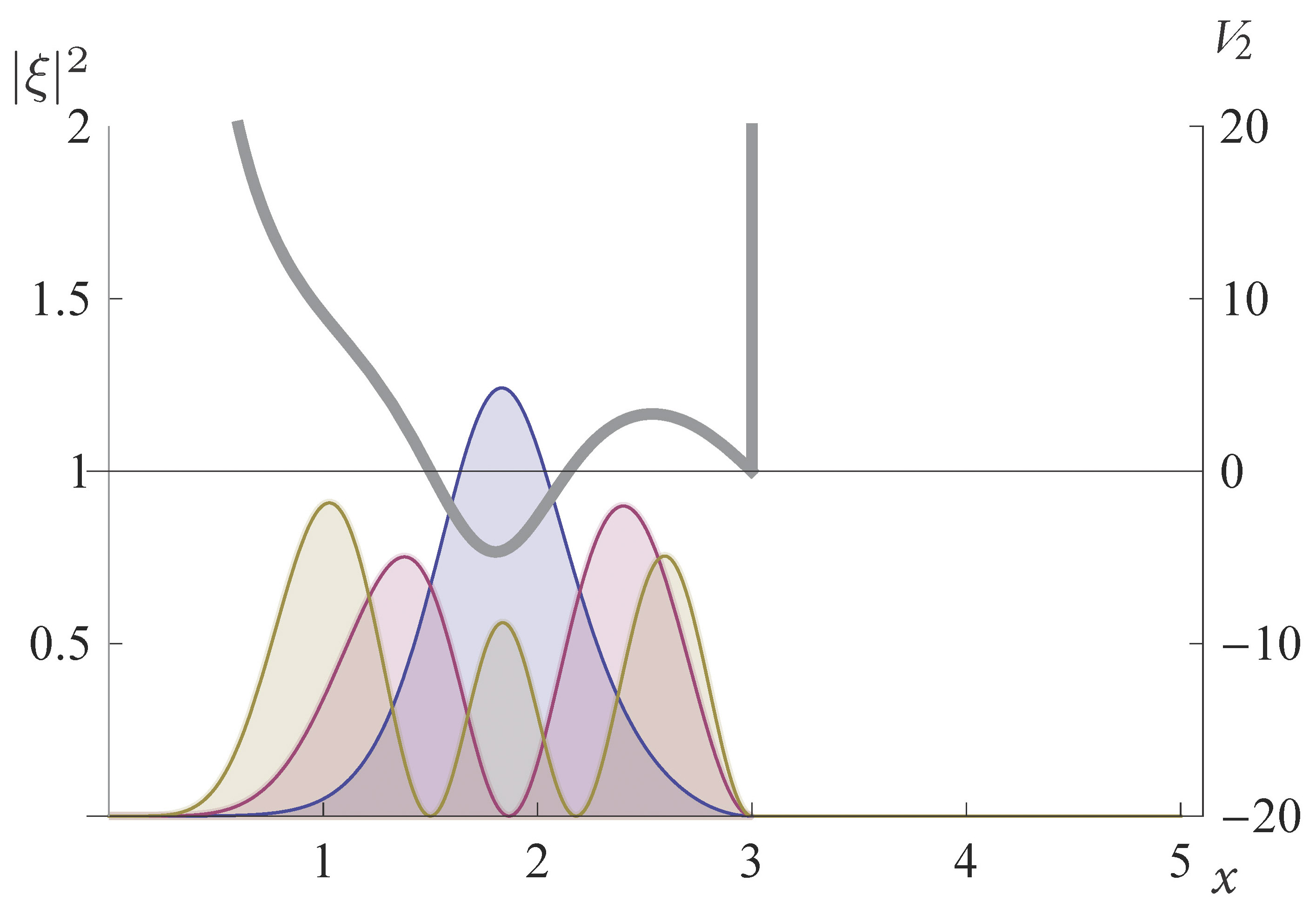}  \\
  \includegraphics[width=.35\textwidth]{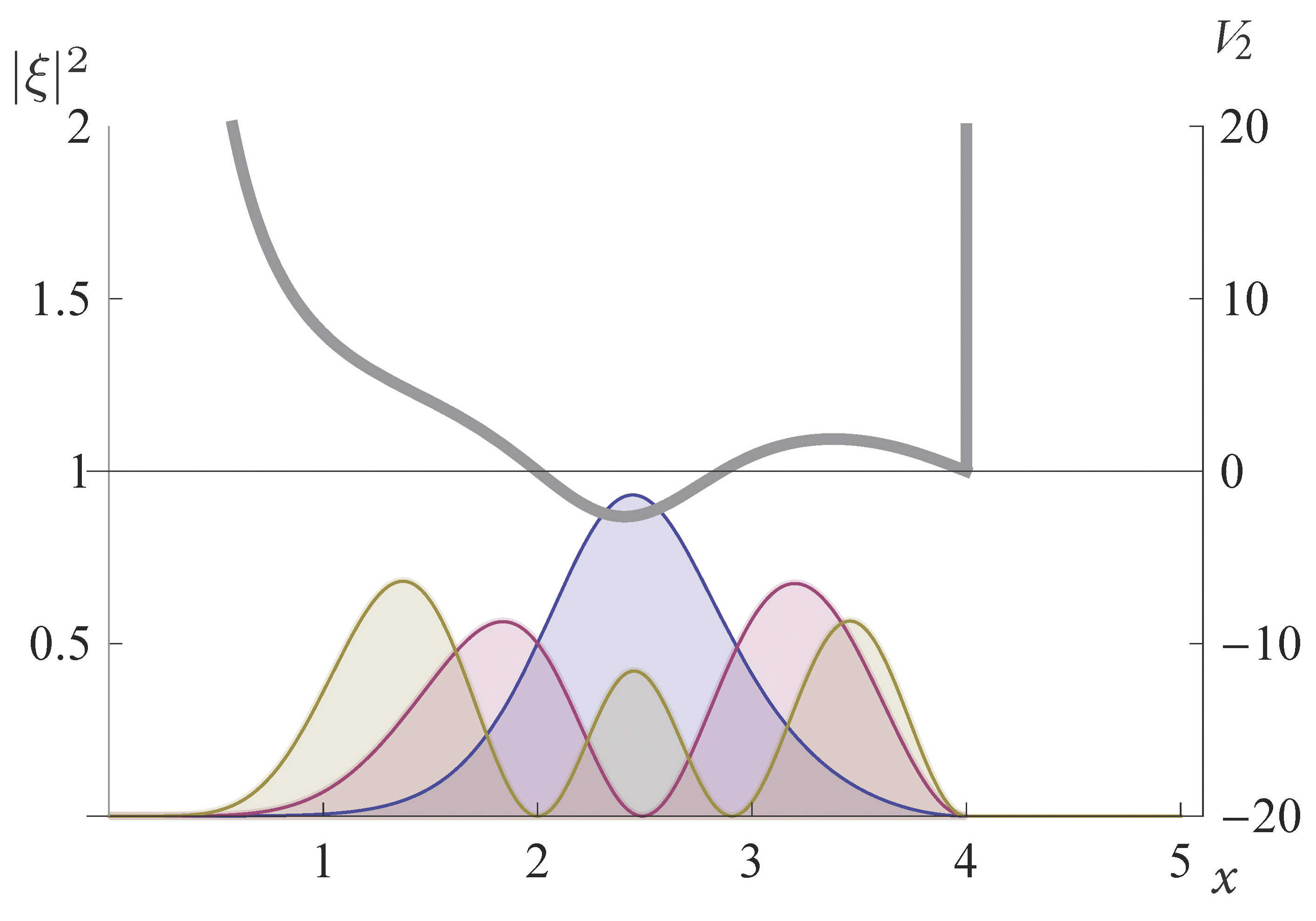}  \qquad
  \includegraphics[width=.35\textwidth]{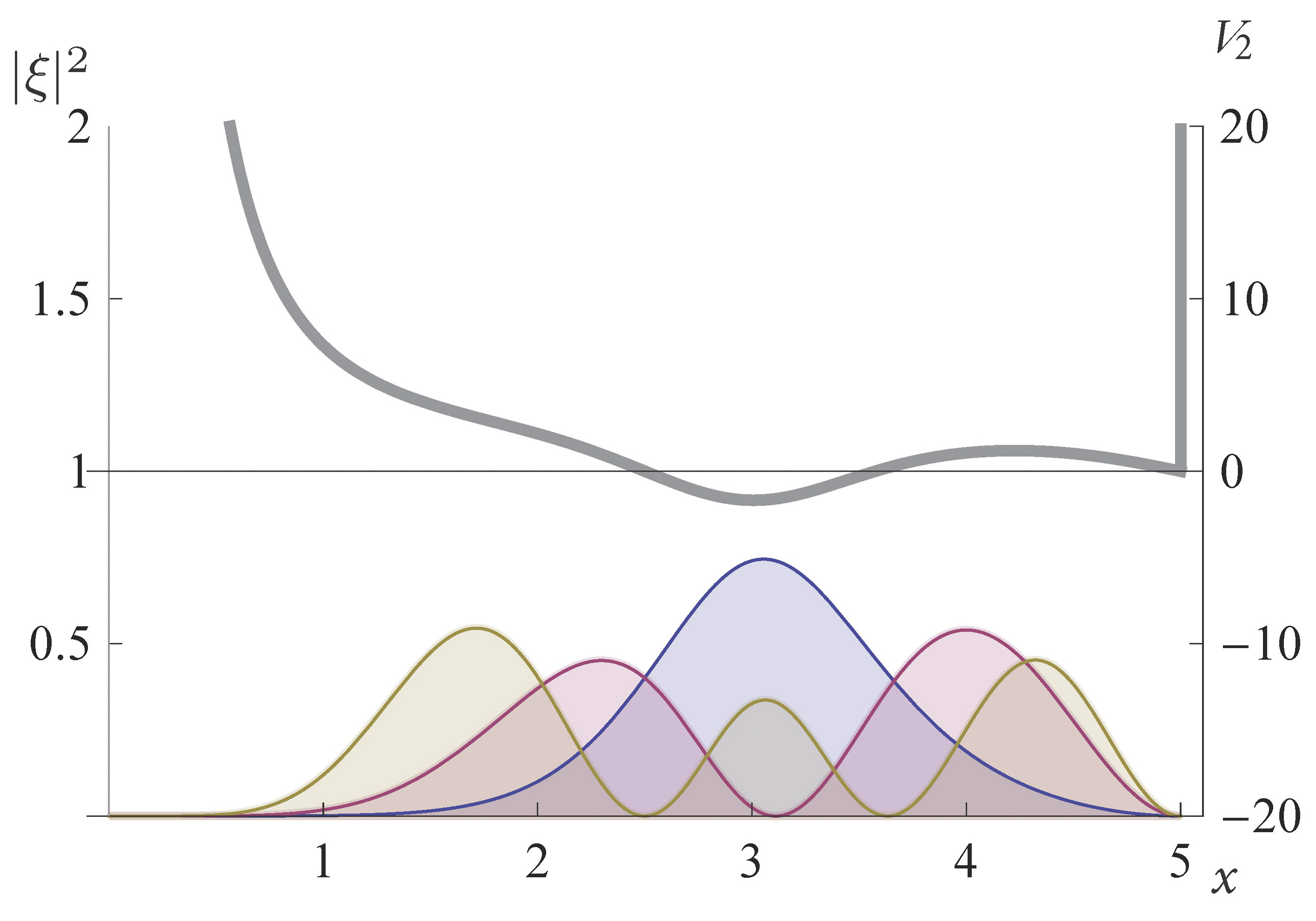}
\caption{In gray a confluent SUSY partner of the infinite square-well potential with a moving barrier, see \eqref{poschl-teller confluente moving}. Moreover,  normalized probability densities are also plotted, see \eqref{xin solutions},  $|\xi_1|^2$ (blue), $|\xi_3|^2$ (purple), $|\xi_4|^2$ (yellow), at four different times: $t=1/4$ (top left), $t=1/2$ (top right), $t=3/4$ (bottom left), $t=1$ (bottom right), for the parameters $L=1$, $\omega=0$ and $m=2$.}
\label{fig:5}      
\end{center}
\end{figure}

Solutions $\xi_n(x,t)$ for these potentials can as well be found with help of intertwining operators $\mathcal{L}_1$ and $\mathcal{L}_2$, see (\ref{square moving sol simplified}) and  (\ref{Poschl operator 2})),  when $n \neq m$:  
\begin{align}
&\xi_n(x,t)= \mathcal{L}_2 \mathcal{L}_1 \phi_n \nonumber \\ & = \left(\frac{\pi }{L} \right)^2 \sqrt{\frac{2}{\ell}} \sin \left( \frac{n \pi}{\ell} x \right) \exp \left\{ i \frac{L}{\ell}\left[ x^2 + \left( \frac{n \pi}{2 L} \right)^2   \right]  \right\}  \nonumber \\
&\times  \vspace{-1cm} \frac{\left[(m^2 +n^2) \sin\left(\frac{2m\pi x}{\ell} \right) + \frac{2 m \pi}{\ell}(m^2 - n^2)(x+\omega \ell)    \right] - 4 m n \ell \cot\left( \frac{n \pi x}{\ell} \right) \sin^2\left( \frac{m \pi x}{\ell} \right)          }{ \frac{2 m \pi}{\ell} (x + \omega \ell) - \sin\left(\frac{2 m \pi x}{\ell} \right)}. \label{xin solutions}
\end{align}
If $n=m$, then the corresponding solution is the missing state $\xi_\epsilon(x,t)$ (see \eqref{missing confluent}):  
\begin{eqnarray}
\xi_\epsilon (x,t)&= &  \sqrt{\frac{2}{\ell}} \sin \left( \frac{m \pi}{\ell} x \right) \nonumber \\ 
& & \times \exp \left\{ i \frac{L}{\ell}\left[ x^2 + \left( \frac{m \pi}{2 L} \right)^2   \right]  \right\} \frac{2 m \pi }{\frac{2m\pi}{\ell}(x+\ell \omega)- \sin\left(\frac{2 m \pi x}{\ell}  \right) }. \label{xi missing}
\end{eqnarray}
Solutions $\xi_n$ satisfy $\lim_{x\to 0} \xi_n(x,t)=\lim_{x\to \ell} \xi_n(x,t)=0$ whereas $\xi_\epsilon$ fulfills
 $\lim_{x\to 0} \xi_\epsilon(x,t)= \lim_{x\to \ell}\xi_\epsilon(x,t)=0$ only if $\omega \neq -1,0$. When $\omega$ is set equal to $-1$ or zero, the function $\xi_\epsilon$ is not square integrable. The confluent SUSY partner of the infinite square-well potential with fixed barriers $\widetilde{V}(y)$, see \eqref{Vy}, are reported in \cite{Fernandez07,Fiset15}, the potentials $V_2(x,t)$ are a dynamic version of them. In Fig. \ref{fig:3} the potential $V_2$ in \eqref{poschl-teller confluente moving} is illustrated, the parameters used where $L=1$, $\omega=0.4$ and $m=2$. This system has sharp edges at $x=0$ and $x=\ell$. Normalized probability densities of three solutions are shown as well, corresponding to $\xi_1$, $\xi_\epsilon$ and $\xi_3$. For the special cases $\omega=-1$ and $\omega=0$ one of the edges of $V_2$ is smooth while the other is sharp, as can be seen in Fig. \ref{fig:4} and Fig. \ref{fig:5}, moreover these special situations does not present a square integrable missing state $\xi_\epsilon$.

\section{Conclusions}

In this article we showed how to generate different infinite potential wells with a moving boundary condition. Through a series of transformations, we obtained the infinite square-well potential, a trigonometric P\"oschl-Teller potential and the confluent SUSY partners of the infinite square-well potential, where one of the barriers is fixed and the other is moving with a constant velocity. For all these systems, exact solutions of the time dependent Schr\"odinger equations fulfilling the moving boundary conditions were given in a closed form.

As a continuation of the present work, it would be interesting to study different sets of coherent and squeezed states for the constructed systems and the calculation of relevant physical quantities and mathematical properties of such states.

\section*{Acknowledgments}
This work has been supported in part by research grants from Natural sciences and engineering research council of Canada (NSERC). ACA would like to thank the Centre de Recherches Math\'ematiques for kind hospitality.

\end{document}